\documentclass[12pt]{article} %,a4paper

\usepackage{physics}
\usepackage{graphicx}
\usepackage{amsmath}
\usepackage{amssymb}
\usepackage{hyperref}
\usepackage[paperwidth=8.5in,paperheight=11in,
  textwidth=6.25in,textheight=9.1in,
  centering]{geometry}

\usepackage{tikz}
\usepackage{cite}
\usepackage{amscd}
\usepackage{setspace}
\usepackage{bbm}
\usepackage{dsfont}
\usepackage{cancel}
\usepackage{relsize}
\usepackage[utf8]{inputenc}
\usepackage{xcolor}
\usepackage{mathtools}
\usepackage{newtxtext,newtxmath}
\numberwithin{equation}{section}

\begin{document}
\begin{titlepage}

\begin{flushright}

\end{flushright}

\vskip 3cm

\begin{center}
{\Large \bf 
On Atiyah-Segal axioms for Witten-type TQFTs }

\vskip 2.0cm

 Wei Gu \\

\bigskip
\begin{tabular}{cc}
Zhejiang Institute of Modern Physics, School of Physics,  Zhejiang University\\
Hangzhou, Zhejiang 310058, China\\

 \end{tabular}

\vskip 1cm

\textbf{Abstract}
\end{center}

 In this paper, we propose a new definition of the trace-map bordism within the Atiyah-Segal framework for Witten-type TQFTs constructed from the topological twist of mass-gapped theories. We demonstrate that these Witten-type TQFTs are unitary under this revised definition and conjecture the self-consistency of the modified bordism category.

\medskip
\noindent

\bigskip
\vfill
\end{titlepage}

\setcounter{tocdepth}{2}
\tableofcontents
\section{Introduction} \label{Intr}
There are two primary ways in which topological quantum field theories arise in physics as the zero-energy (topological) sector of more general quantum field theories (QFTs): the far-infrared (IR) limits of gapped theories \cite{Schwarz:1978cn,Witten:1988hf} and the topological twisting of supersymmetric (SUSY) theories \cite{Witten:1988ze, Witten:1988xj}. Inspired by Witten's work, Atiyah \cite{Atiyah:1989vu} adapted Segal's framework \cite{Segal:2002ei} — originally a categorical foundation for conformal field theories — to topological theories.  \footnote{There are some excellent reviews on this subject written by mathematicians \cite{Freed:1992TQFT, Teleman:2016jhd}. However, the author of this paper learned most of this language from Moore's lecture note \cite{Moore:2015TQFT} and his paper with Segal \cite{Moore:2006dw}.}

TQFTs provide a deeply mutually enriching interplay between physics and mathematics. However, there are notable differences between the two frameworks. In the physical formulation, a TQFT is expected to assign Hilbert spaces and transition amplitudes to topological manifolds. The Hermitian structure of the TQFT state space may be inherited from the underlying quantum field theory in which the TQFT is embedded. In contrast, within the Atiyah-Segal axiomatic framework, the state space is fundamentally treated as a plain vector space without an inherent Hermitian structure. We will review the basic setup of the Atiyah-Segal axioms in Section \ref{AS:axioms} following Moore's presentation in \cite{Moore:2015TQFT}. We will see that in a unitary TQFT, the Hermitian inner product arises from gluing a state and its dual via the bended cylinder bordism, yielding their transition amplitude through annihilation into the vacuum (empty space). From a physics perspective, this inner product may be encoded in two-point correlation functions of the TQFT. Mathematically, in the Atiyah-Segal framework, the inner product is constructed through the composition of the multiplication structure with a trace map. 

For a TQFT derived from the infrared (IR) sector of a mass-gapped QFT — commonly referred to as Schwarz-type TQFTs, including examples such as the gauged Wess-Zumino-Witten (GWZW) models, Chern-Simons gauge theories, and so on — the Atiyah-Segal framework provides a perfect mathematical foundation for these physical theories.  However, when naively applying the Atiyah-Segal framework to generic Witten-type TQFTs (as will be discussed in Sections \ref{AS:axioms} and \ref{BWTQFT}), one finds that the inner product of the ground state is generally not positive-definite. If we insist on interpreting this vector space as a Hilbert space, the resulting non-positive-definite structure would imply that such Witten-type TQFTs may violate unitarity. In the two-dimensional case, however, this conclusion would contradict the correspondence established in Witten's seminal work \cite{Witten:1993xi}:
\begin{center}
    Quantum cohomology of $Gr(k,n)$ $\cong$ Verlinde algebra of $U(k)_{n-k}$,
\end{center}
as we cannot naturally reconcile a non-unitary theory with a unitary one (the right-hand side of this correspondence). Beyond this issue, additional inconsistencies arise when naively applying the current framework to Witten-type TQFTs. As demonstrated in Section 3, the composition of co-multiplication with the trace fails to yield the identity functor — equivalent to the time evolution operator ($e^{iHt}=1$) — in TQFT. This mathematical discrepancy stems physically from either:
(i) the axial symmetry anomaly in the topological A-model, or
(ii) the vector R-symmetry anomaly in its mirror dual  \cite{Hori:2000kt, Gu:2018fpm}. 

In this paper, we address the following fundamental question in two dimensions:
\begin{center}
   \textit{Can the Atiyah-Segal framework be systematically extended to consistently describe Witten-type TQFTs? } 
\end{center}
In Section \ref{BWTQFT}, we demonstrate that the only required modification concerns the trace map and its associated bordism:
\begin{equation*}
    \tau({\cal O}) \Rightarrow \tau({\cal O}\cdot {\color{red}\bullet} ),
\end{equation*}
where ${\cal O}$ denotes a general state while $\color{red}\bullet$ represents a ``background particle" carrying the opposite axial charge of the theory. This modification is motivated by the correspondence in correlation functions. For example, in the case of $\mathbb{CP}^n$ studied in \cite{Witten:1993xi}, the correspondence is $\langle {\cal O} \cdot \sigma^{n(1-g)}\rangle_{\mathbb{CP}^n}=\langle {\cal O} \rangle_{U(1)_n}$ on the genus-$g$ spacetime. This phenomenon can be alternatively understood through the constraint of vanishing net charges on compact surfaces, as required by the Gauss law. The topological insertion serves to maintain this condition while preserving the surface topology.
However, an important subtlety arises: while inner products are conventionally defined via two-point correlators, the modified case appears to encode them in three-point correlators. What explains this apparent discrepancy? The resolution becomes clearer upon examining a three-dimensional theory \cite{Kapustin:2013hpk}, where background particle insertions emerge naturally from the Lagrangian of the KK-reduced 3D theory. This higher-dimensional perspective reveals that: the inner product retains its standard interpretation as a two-point correlator, and the topology of the modified trace map for the identity operator corresponds to an effective two-sphere. Additional details supporting this interpretation are developed in\cite[Section 5]{Gu:2025gtb}. 

In this extended framework, a functor can be naturally constructed from the Witten-type TQFT to its corresponding Schwarz-type counterpart, suggesting their equivalence in the bordism language. However, we show in Section \ref{SBTQFT} that they differ in their summation over bordisms.

We conclude this section by noting that our discussion is limited to closed strings, leaving a detailed analysis of the open string case \cite{Moore:2006dw} for future investigation. Furthermore, our work focuses exclusively on Witten-type TQFTs constructed from mass-gapped theories. For topological superconformal field theories with a positive central charge, our approach may be insufficient to establish unitarity. It is well-established that coupling to gravity is necessary to incorporate higher-genus amplitudes.

\section{The Atiyah-Segal axioms for two-dimensional TQFTs}\label{AS:axioms}
In this section, we briefly review the Atiyah-Segal axioms for Schwarz-type TQFTs and establish our notation. A D-dimensional TQFT is defined as a symmetric monoidal functor:
\begin{equation}\label{TQFT:Funt}
Z: \left(\rm{Bord}_{\langle D-1,D \rangle}, \coprod\right) \mapsto \left(\rm{Vect_{\kappa}}, \otimes\right),
\end{equation}
where $\rm{Bord}_{\langle D-1,D \rangle}$ is a tensor category whose objects are compact-oriented smooth (D-1)-dimensional manifolds without boundary \footnote{The category of boundary condition in the case that the closed string algebra is semi-simple has been studied in \cite{Moore:2006dw}. See \cite{Kapustin:2003ga, Herbst:2004ax} for concrete models.}\footnote{In the mathematical literature (e.g.,\cite{Lurie:2008TQFT}), the notion of \textit{extended TQFT} has been introduced, which enriches the standard bordism category by incorporating additional algebraic structures.}, and the morphisms are oriented D-dimensional bordisms (considered up to boundary-fixing diffeomorphisms); while Vect$_{\kappa}$ denotes the category of finite-dimensional vector spaces over a field $\kappa$ (typically chosen as $\mathbb{C}$ in physical applications). The quantum nature of the theory manifests itself through the functor's mapping of disjoint union ($\coprod$) on the left-hand side of Eq.(\ref{TQFT:Funt}) to the tensor product ($\otimes$) on the right-hand side.

 This paper focuses primarily on the two-dimensional case (D=2), where the generating objects are circles modulo orientation-reversing diffeomorphisms (reflections). A fundamental theorem in this field establishes the equivalence between 2D oriented TQFTs and Frobenius algebras. Let $\cal C$ denote a \textit{ commutative Frobenius algebra} (whose precise definition will be given later), this correspondence implies the existence of a homomorphism $Z: \left(\rm{Bord}_{\langle 1,2 \rangle}, \coprod\right) \mapsto \left(\rm{Vect_{\mathbb{C}}}, \otimes\right)$ with $Z(S^1)=\cal C$.  The physical interpretation of this result might be first suggested by Dijkgraaf in his thesis \cite{Dijkgraaf:1989CFT}, with subsequent rigorous mathematical proofs for TQFTs appearing in \cite{Abrams:1996TQFT}. A detailed proof along with generalizations can also be found in the appendix of \cite{Moore:2006dw}. 
 
Physical processes involving amplitudes with $p$ incoming and $q$ outgoing ``point particles" naturally described in this framework. In two dimensions, the circle $S^1$ serves as the link of a point, which implies that the vector space ${\cal C}=Z(S^1)$ corresponds to the state space of point particles (or point operators) in the TQFT. Consequently, $n$ incoming point particles can be mathematically represented by the following functor:
\begin{equation}\label{Incoming:Funt}
Z\left(S^1\coprod \cdots\coprod S^1\right) = {\cal C}^{\otimes p}.
\end{equation}
Furthermore, the transition amplitudes can be expressed as a linear map
\begin{equation}\label{Tran:Funt}
Z\left(\Sigma\right):  {\cal C}^{\otimes p}\xrightarrow \quad{\cal C}^{\otimes q},
\end{equation}
where $\Sigma$ is a Riemann surface (representing the space-time evolution) with $p$ incoming and $q$ outgoing particle boundary components. 

To properly encode the additional structure of the vector spaces, we must carefully analyze the topology of the corresponding bordisms. The key insight is that the bordism category can be presented in terms of generators and relations. In two dimensions, the elementary bordisms (generators) consist of: the multiplication bordism (pair of pants creating a junction) and co-multiplication structures are

\begin{center}
\tikzset{every picture/.style={line width=0.75pt}} %set default line width to 0.75pt        

\begin{tikzpicture}[x=0.75pt,y=0.75pt,yscale=-1,xscale=1]

%Flowchart: Connector [id:dp3642784396378971] 
\draw   (3.5,120) .. controls (3.5,111.72) and (8.09,105) .. (13.75,105) .. controls (19.41,105) and (24,111.72) .. (24,120) .. controls (24,128.28) and (19.41,135) .. (13.75,135) .. controls (8.09,135) and (3.5,128.28) .. (3.5,120) -- cycle ;
%Flowchart: Connector [id:dp45394820982181416] 
\draw   (4.5,178) .. controls (4.5,169.72) and (9.09,163) .. (14.75,163) .. controls (20.41,163) and (25,169.72) .. (25,178) .. controls (25,186.28) and (20.41,193) .. (14.75,193) .. controls (9.09,193) and (4.5,186.28) .. (4.5,178) -- cycle ;
%Flowchart: Connector [id:dp30048438073738004] 
\draw   (69.5,148) .. controls (69.5,139.72) and (74.09,133) .. (79.75,133) .. controls (85.41,133) and (90,139.72) .. (90,148) .. controls (90,156.28) and (85.41,163) .. (79.75,163) .. controls (74.09,163) and (69.5,156.28) .. (69.5,148) -- cycle ;
%Curve Lines [id:da1297872933913239] 
\draw    (13.75,105) .. controls (55.5,105) and (39.5,137) .. (79.75,133) ;
%Curve Lines [id:da3236195129249373] 
\draw    (14.75,193) .. controls (53.5,181) and (39.75,193) .. (79.75,163) ;
%Curve Lines [id:da763554013020583] 
\draw    (13.75,135) .. controls (60.5,123) and (63.5,168) .. (14.75,163) ;
%Flowchart: Connector [id:dp43860113975266135] 
\draw   (247.78,176.32) .. controls (247.72,184.61) and (243.08,191.29) .. (237.42,191.25) .. controls (231.76,191.2) and (227.22,184.45) .. (227.28,176.17) .. controls (227.34,167.89) and (231.98,161.2) .. (237.64,161.25) .. controls (243.3,161.29) and (247.84,168.04) .. (247.78,176.32) -- cycle ;
%Flowchart: Connector [id:dp9808788454475847] 
\draw   (247.22,118.32) .. controls (247.15,126.6) and (242.51,133.28) .. (236.85,133.24) .. controls (231.19,133.2) and (226.65,126.45) .. (226.72,118.16) .. controls (226.78,109.88) and (231.42,103.2) .. (237.08,103.24) .. controls (242.74,103.28) and (247.28,110.03) .. (247.22,118.32) -- cycle ;
%Flowchart: Connector [id:dp2587132199724371] 
\draw   (181.99,147.83) .. controls (181.93,156.11) and (177.29,162.79) .. (171.63,162.75) .. controls (165.97,162.71) and (161.43,155.96) .. (161.49,147.68) .. controls (161.56,139.39) and (166.2,132.71) .. (171.86,132.75) .. controls (177.52,132.8) and (182.06,139.55) .. (181.99,147.83) -- cycle ;
%Curve Lines [id:da781796031170328] 
\draw    (237.42,191.25) .. controls (198.77,177.96) and (211.87,164.05) .. (171.63,162.75) ;
%Curve Lines [id:da051850977991430014] 
\draw    (237.08,103.24) .. controls (194.5,110) and (212.08,103.05) .. (171.86,132.75) ;
%Curve Lines [id:da01609465548874922] 
\draw    (237.64,161.25) .. controls (190.81,172.9) and (188.14,127.87) .. (236.85,133.24) ;

% Text Node
\draw (12,212.4) node [anchor=north west][inner sep=0.75pt]   [xscale=1,yscale=1]  {${\textit{m}}: {{\cal C}\otimes{\cal C}\rightarrow {\cal C}}$};
% Text Node
\draw (167,212.4) node [anchor=north west][inner sep=0.75pt]   [xscale=1,yscale=1]  {${\Delta}: {\cal C}\rightarrow {\cal C}\otimes{\cal C}$};

\end{tikzpicture},
\end{center}
the unit map and the co-unit map (also called the trace map) are
\begin{center}

\tikzset{every picture/.style={line width=0.75pt}} %set default line width to 0.75pt        

\begin{tikzpicture}[x=0.75pt,y=0.75pt,yscale=-1,xscale=1]
%uncomment if require: \path (0,176); %set diagram left start at 0, and has height of 176

%Flowchart: Connector [id:dp4507729894860384] 
\draw   (37,95.2) .. controls (37,80.29) and (44.61,68.2) .. (54,68.2) .. controls (63.39,68.2) and (71,80.29) .. (71,95.2) .. controls (71,110.11) and (63.39,122.2) .. (54,122.2) .. controls (44.61,122.2) and (37,110.11) .. (37,95.2) -- cycle ;
%Curve Lines [id:da7232271355427081] 
\draw    (51.6,68) .. controls (-29,75) and (7.8,122.2) .. (54,122.2) ;
%Flowchart: Connector [id:dp3760466491705351] 
\draw   (199.36,96.18) .. controls (199.74,111.09) and (192.45,123.37) .. (183.06,123.62) .. controls (173.68,123.86) and (165.76,111.98) .. (165.37,97.07) .. controls (164.98,82.16) and (172.27,69.88) .. (181.66,69.64) .. controls (191.05,69.39) and (198.97,81.28) .. (199.36,96.18) -- cycle ;
%Curve Lines [id:da8168592751480289] 
\draw    (185.47,123.76) .. controls (265.86,114.66) and (227.84,68.43) .. (181.66,69.64) ;

% Text Node
\draw (10.2,134.8) node [anchor=north west][inner sep=0.75pt]  [xscale=1,yscale=1]  {$\rm{I}: {\mathbb C}\rightarrow {\cal C}$};
% Text Node
\draw (170.2,136.2) node [anchor=north west][inner sep=0.75pt]  [xscale=1,yscale=1] 
{$\tau: {\cal C}\rightarrow {\mathbb C}$ };
% Text Node
\draw (28,157.4) node [anchor=north west][inner sep=0.75pt]  [xscale=1,yscale=1]  {$1\mapsto 1_{\cal C}$};

\end{tikzpicture},
\end{center}
respectively. All other bordisms in this category can be decomposed into compositions of these fundamental morphisms. As a key example, the identity functor admits a canonical factorization through the unit and multiplication morphisms via the following diagrammatic relation:

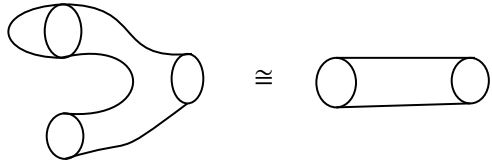
\begin{center}

\tikzset{every picture/.style={line width=0.75pt}} %set default line width to 0.75pt        

\begin{tikzpicture}[x=0.75pt,y=0.75pt,yscale=-1,xscale=1]

%Flowchart: Connector [id:dp3642784396378971] 
\draw   (23.5,124.5) .. controls (23.5,117.04) and (27.64,111) .. (32.75,111) .. controls (37.86,111) and (42,117.04) .. (42,124.5) .. controls (42,131.96) and (37.86,138) .. (32.75,138) .. controls (27.64,138) and (23.5,131.96) .. (23.5,124.5) -- cycle ;
%Flowchart: Connector [id:dp45394820982181416] 
\draw   (24.5,177.5) .. controls (24.5,171.15) and (28.64,166) .. (33.75,166) .. controls (38.86,166) and (43,171.15) .. (43,177.5) .. controls (43,183.85) and (38.86,189) .. (33.75,189) .. controls (28.64,189) and (24.5,183.85) .. (24.5,177.5) -- cycle ;
%Flowchart: Connector [id:dp30048438073738004] 
\draw   (87.5,148.5) .. controls (87.5,141.6) and (91.08,136) .. (95.5,136) .. controls (99.92,136) and (103.5,141.6) .. (103.5,148.5) .. controls (103.5,155.4) and (99.92,161) .. (95.5,161) .. controls (91.08,161) and (87.5,155.4) .. (87.5,148.5) -- cycle ;
%Curve Lines [id:da1297872933913239] 
\draw    (32.75,111) .. controls (74.5,111) and (57.5,140) .. (97.75,136) ;
%Curve Lines [id:da3236195129249373] 
\draw    (33.75,189) .. controls (72.5,177) and (55.5,191) .. (95.5,161) ;
%Curve Lines [id:da763554013020583] 
\draw    (31.75,138) .. controls (78.5,126) and (81.5,171) .. (32.75,166) ;
%Flowchart: Connector [id:dp8471479756212953] 
\draw   (160.5,150.5) .. controls (160.5,143.6) and (164.98,138) .. (170.5,138) .. controls (176.02,138) and (180.5,143.6) .. (180.5,150.5) .. controls (180.5,157.4) and (176.02,163) .. (170.5,163) .. controls (164.98,163) and (160.5,157.4) .. (160.5,150.5) -- cycle ;
%Straight Lines [id:da06231118546084191] 
\draw    (170.5,138) -- (240.5,138) ;
%Straight Lines [id:da08760758549305958] 
\draw    (170.5,163) -- (238.13,161) ;
%Flowchart: Connector [id:dp061766882323054606] 
\draw   (228.75,149.5) .. controls (228.75,143.15) and (232.95,138) .. (238.13,138) .. controls (243.3,138) and (247.5,143.15) .. (247.5,149.5) .. controls (247.5,155.85) and (243.3,161) .. (238.13,161) .. controls (232.95,161) and (228.75,155.85) .. (228.75,149.5) -- cycle ;
%Curve Lines [id:da8388334680251182] 
\draw    (32.75,111) .. controls (-5.5,113) and (-2.5,138) .. (32.75,138) ;

% Text Node
\draw (34,213.4) node [anchor=north west][inner sep=0.75pt]  [xscale=1,yscale=1]  {$1_{\cal C}\cdot \cal C=\cal C $};
% Text Node
\draw (127,142.4) node [anchor=north west][inner sep=0.75pt]   [xscale=1,yscale=1]  {$\cong$};

\end{tikzpicture}.
\end{center}

\noindent{}The identity functor's operator manifests its topological character through the time evolution operator $U(t)=e^{iHt}$.
When the Hamiltonian $H$ vanishes, this reduces to the identity operation. More generally in topological theories, the Hamiltonian is BRST-exact, which likewise ensures trivial time evolution. Two additional important bordisms 
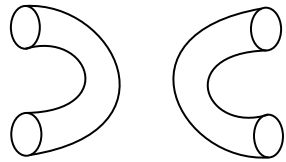
\begin{center}
\tikzset{every picture/.style={line width=0.75pt}} %set default line width to 0.75pt        
\begin{tikzpicture}[x=0.75pt,y=0.75pt,yscale=-1,xscale=1]
%uncomment if require: \path (0,176); %set diagram left start at 0, and has height of 176

%Flowchart: Connector [id:dp5255615149000737] 
\draw   (2,56.6) .. controls (2,50.64) and (5.27,45.8) .. (9.3,45.8) .. controls (13.33,45.8) and (16.6,50.64) .. (16.6,56.6) .. controls (16.6,62.56) and (13.33,67.4) .. (9.3,67.4) .. controls (5.27,67.4) and (2,62.56) .. (2,56.6) -- cycle ;
%Flowchart: Connector [id:dp2722199131110099] 
\draw   (2.2,110.6) .. controls (2.2,104.64) and (5.6,99.8) .. (9.8,99.8) .. controls (14,99.8) and (17.4,104.64) .. (17.4,110.6) .. controls (17.4,116.56) and (14,121.4) .. (9.8,121.4) .. controls (5.6,121.4) and (2.2,116.56) .. (2.2,110.6) -- cycle ;
%Curve Lines [id:da5522128140043935] 
\draw    (9.3,45.8) .. controls (54.2,43.4) and (87.8,109) .. (11.8,121) ;
%Curve Lines [id:da711579998404773] 
\draw    (9.3,67.4) .. controls (41.4,57.8) and (57,99) .. (9.8,99.8) ;
%Flowchart: Connector [id:dp8608078538348912] 
\draw   (139.23,111.36) .. controls (139.31,117.32) and (136.1,122.2) .. (132.07,122.26) .. controls (128.04,122.31) and (124.71,117.52) .. (124.63,111.56) .. controls (124.55,105.59) and (127.75,100.72) .. (131.78,100.66) .. controls (135.81,100.61) and (139.14,105.4) .. (139.23,111.36) -- cycle ;
%Flowchart: Connector [id:dp2703800369134939] 
\draw   (138.29,57.37) .. controls (138.37,63.33) and (135.03,68.21) .. (130.84,68.27) .. controls (126.64,68.33) and (123.17,63.54) .. (123.09,57.57) .. controls (123.01,51.61) and (126.35,46.73) .. (130.54,46.67) .. controls (134.74,46.61) and (138.21,51.4) .. (138.29,57.37) -- cycle ;
%Curve Lines [id:da23599691921891197] 
\draw    (132.07,122.26) .. controls (87.21,125.27) and (52.72,60.13) .. (128.55,47.1) ;
%Curve Lines [id:da08875303710615501] 
\draw    (131.78,100.66) .. controls (99.81,110.7) and (83.65,69.71) .. (130.84,68.27) ;

\end{tikzpicture}
\end{center}
 through the compositions: $\tau\circ{\textit{m}}$ : ${\cal C}\otimes{\cal C}\rightarrow {\mathbb C}$ and $\Delta\circ{\rm{I}}$ : ${\mathbb C}\rightarrow {\cal C}\otimes{\cal C}$, respectively. Then from the S-diagram
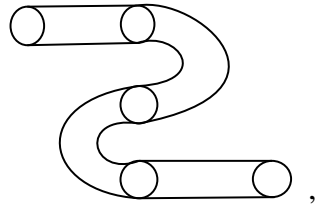
\begin{center}

\tikzset{every picture/.style={line width=0.75pt}} %set default line width to 0.75pt        

\begin{tikzpicture}[x=0.75pt,y=0.75pt,yscale=-1,xscale=1]

%Flowchart: Connector [id:dp8940861224340374] 
\draw   (2,86.52) .. controls (2,81.14) and (5.79,76.79) .. (10.47,76.79) .. controls (15.15,76.79) and (18.94,81.14) .. (18.94,86.52) .. controls (18.94,91.89) and (15.15,96.25) .. (10.47,96.25) .. controls (5.79,96.25) and (2,91.89) .. (2,86.52) -- cycle ;
%Straight Lines [id:da2248139295394913] 
\draw    (10.47,76.79) -- (66.11,76.22) ;
%Straight Lines [id:da522827692920293] 
\draw    (10.47,96.25) -- (66.11,95.1) ;
%Flowchart: Connector [id:dp0064580656130279745] 
\draw   (57.64,85.66) .. controls (57.64,80.44) and (61.43,76.22) .. (66.11,76.22) .. controls (70.78,76.22) and (74.58,80.44) .. (74.58,85.66) .. controls (74.58,90.87) and (70.78,95.1) .. (66.11,95.1) .. controls (61.43,95.1) and (57.64,90.87) .. (57.64,85.66) -- cycle ;
%Flowchart: Connector [id:dp7798313477583939] 
\draw   (57.47,126.29) .. controls (57.47,121.07) and (61.63,116.85) .. (66.77,116.85) .. controls (71.91,116.85) and (76.07,121.07) .. (76.07,126.29) .. controls (76.07,131.5) and (71.91,135.73) .. (66.77,135.73) .. controls (61.63,135.73) and (57.47,131.5) .. (57.47,126.29) -- cycle ;
%Flowchart: Connector [id:dp6439978229001175] 
\draw   (57.47,164.06) .. controls (57.47,158.84) and (61.63,154.62) .. (66.77,154.62) .. controls (71.91,154.62) and (76.07,158.84) .. (76.07,164.06) .. controls (76.07,169.27) and (71.91,173.5) .. (66.77,173.5) .. controls (61.63,173.5) and (57.47,169.27) .. (57.47,164.06) -- cycle ;
%Curve Lines [id:da2685188353318506] 
\draw    (66.11,76.22) .. controls (96.67,71.07) and (151.8,119.71) .. (66.77,135.73) ;
%Curve Lines [id:da030974940449317923] 
\draw    (66.11,95.1) .. controls (86.7,88.23) and (104.64,114.56) .. (66.77,116.85) ;
%Curve Lines [id:da48433807071463875] 
\draw    (66.77,116.85) .. controls (9.64,124.29) and (17.61,170.64) .. (66.77,173.5) ;
%Curve Lines [id:da6204323103807131] 
\draw    (66.77,135.73) .. controls (35.55,131.15) and (42.19,163.77) .. (66.77,154.62) ;
%Straight Lines [id:da3499529527671469] 
\draw    (66.77,154.62) -- (133.87,154.62) ;
%Straight Lines [id:da3903316603872573] 
\draw    (66.77,173.5) -- (135.2,172.93) ;
%Flowchart: Connector [id:dp07177507759389024] 
\draw   (124.23,163.77) .. controls (124.23,158.71) and (128.55,154.62) .. (133.87,154.62) .. controls (139.19,154.62) and (143.5,158.71) .. (143.5,163.77) .. controls (143.5,168.83) and (139.19,172.93) .. (133.87,172.93) .. controls (128.55,172.93) and (124.23,168.83) .. (124.23,163.77) -- cycle ;
\end{tikzpicture}, 
\end{center}
one can straightforwardly verify that the quadratic form 
\begin{equation}\label{Qud:Funt}
Q(\sigma^i, \sigma^j)=\tau(\sigma^i\sigma^j)=Q^{ij}, \quad \sigma^i\in \cal C
\end{equation}
is non-degenerate, as required by the following condition:
\begin{equation}\label{Qud2:Funt}
\sum_j \Delta_{ij}Q^{jk}=\delta^k_i.
\end{equation}
 Then, combining this with the associativity 
 
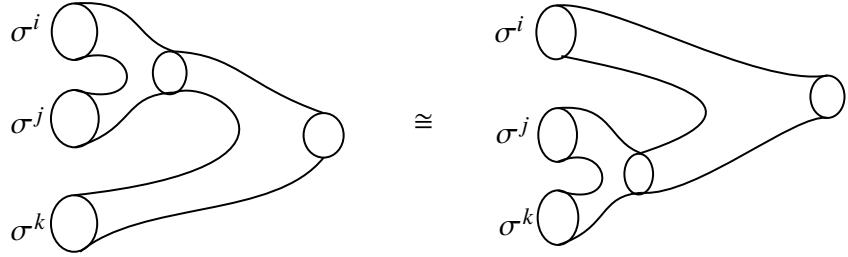
\begin{center}
\tikzset{every picture/.style={line width=0.75pt}} %set default line width to 0.75pt        

\begin{tikzpicture}[x=0.75pt,y=0.75pt,yscale=-1,xscale=1]

%Flowchart: Connector [id:dp09460067594428812] 
\draw   (30.67,93.48) .. controls (30.67,85.61) and (35.8,79.22) .. (42.13,79.22) .. controls (48.46,79.22) and (53.6,85.61) .. (53.6,93.48) .. controls (53.6,101.36) and (48.46,107.74) .. (42.13,107.74) .. controls (35.8,107.74) and (30.67,101.36) .. (30.67,93.48) -- cycle ;
%Flowchart: Connector [id:dp9352946377455752] 
\draw   (30.28,137.35) .. controls (30.28,129.35) and (35.68,122.87) .. (42.33,122.87) .. controls (48.98,122.87) and (54.37,129.35) .. (54.37,137.35) .. controls (54.37,145.34) and (48.98,151.82) .. (42.33,151.82) .. controls (35.68,151.82) and (30.28,145.34) .. (30.28,137.35) -- cycle ;
%Flowchart: Connector [id:dp9647383308847671] 
\draw   (30.28,190.28) .. controls (30.28,182.41) and (35.5,176.02) .. (41.94,176.02) .. controls (48.38,176.02) and (53.6,182.41) .. (53.6,190.28) .. controls (53.6,198.16) and (48.38,204.54) .. (41.94,204.54) .. controls (35.5,204.54) and (30.28,198.16) .. (30.28,190.28) -- cycle ;
%Curve Lines [id:da3257302731616143] 
\draw    (42.13,79.22) .. controls (75.36,76.63) and (71.47,97.37) .. (92.45,103.42) ;
%Curve Lines [id:da7938708194273514] 
\draw    (42.13,107.74) .. controls (73.8,95.64) and (80.8,134.54) .. (42.33,122.87) ;
%Curve Lines [id:da9745738928521192] 
\draw    (42.33,151.82) .. controls (73.02,143.18) and (66.03,125.03) .. (92.45,124.17) ;
%Flowchart: Connector [id:dp9305261954044054] 
\draw   (81.67,114.23) .. controls (81.67,108.26) and (85.47,103.42) .. (90.17,103.42) .. controls (94.86,103.42) and (98.67,108.26) .. (98.67,114.23) .. controls (98.67,120.19) and (94.86,125.03) .. (90.17,125.03) .. controls (85.47,125.03) and (81.67,120.19) .. (81.67,114.23) -- cycle ;
%Curve Lines [id:da5366399631448724] 
\draw    (92.45,103.42) .. controls (124.31,105.15) and (128.2,120.71) .. (167.78,134.54) ;
%Curve Lines [id:da09118592499597977] 
\draw    (45,204.54) .. controls (76.08,178.62) and (142.92,188.12) .. (167.78,157.01) ;
%Shape: Ellipse [id:dp9335024343794197] 
\draw   (157.68,145.77) .. controls (157.68,139.57) and (162.2,134.54) .. (167.78,134.54) .. controls (173.36,134.54) and (177.89,139.57) .. (177.89,145.77) .. controls (177.89,151.98) and (173.36,157.01) .. (167.78,157.01) .. controls (162.2,157.01) and (157.68,151.98) .. (157.68,145.77) -- cycle ;
%Flowchart: Connector [id:dp3290988201985736] 
\draw   (276.07,145.61) .. controls (276.07,138.12) and (280.44,132.04) .. (285.83,132.04) .. controls (291.22,132.04) and (295.58,138.12) .. (295.58,145.61) .. controls (295.58,153.1) and (291.22,159.17) .. (285.83,159.17) .. controls (280.44,159.17) and (276.07,153.1) .. (276.07,145.61) -- cycle ;
%Flowchart: Connector [id:dp14218599578160807] 
\draw   (275.74,187.32) .. controls (275.74,179.72) and (280.33,173.55) .. (285.99,173.55) .. controls (291.65,173.55) and (296.25,179.72) .. (296.25,187.32) .. controls (296.25,194.92) and (291.65,201.09) .. (285.99,201.09) .. controls (280.33,201.09) and (275.74,194.92) .. (275.74,187.32) -- cycle ;
%Flowchart: Connector [id:dp9736850900131477] 
\draw   (275.08,93.82) .. controls (275.08,86.33) and (279.52,80.26) .. (285,80.26) .. controls (290.48,80.26) and (294.92,86.33) .. (294.92,93.82) .. controls (294.92,101.31) and (290.48,107.39) .. (285,107.39) .. controls (279.52,107.39) and (275.08,101.31) .. (275.08,93.82) -- cycle ;
%Curve Lines [id:da19113039122138742] 
\draw    (285.83,132.04) .. controls (314.11,129.58) and (310.8,149.3) .. (328.66,155.06) ;
%Curve Lines [id:da46509882749220866] 
\draw    (285.83,159.17) .. controls (312.78,147.66) and (318.74,184.65) .. (285.99,173.55) ;
%Curve Lines [id:da1939473703313841] 
\draw    (285.99,201.09) .. controls (312.12,192.87) and (306.17,175.61) .. (328.66,174.78) ;
%Flowchart: Connector [id:dp11673538169317821] 
\draw   (319.48,165.33) .. controls (319.48,159.66) and (322.72,155.06) .. (326.72,155.06) .. controls (330.71,155.06) and (333.95,159.66) .. (333.95,165.33) .. controls (333.95,171.01) and (330.71,175.61) .. (326.72,175.61) .. controls (322.72,175.61) and (319.48,171.01) .. (319.48,165.33) -- cycle ;
%Curve Lines [id:da9561635071948004] 
\draw    (328.66,174.78) .. controls (355.79,176.43) and (406.02,134.51) .. (421.9,136.98) ;
%Curve Lines [id:da9778083653448956] 
\draw    (287.69,105.74) .. controls (331.97,114.78) and (403.42,130.4) .. (326.72,155.06) ;
%Curve Lines [id:da47967916184779846] 
\draw    (285,80.26) .. controls (312.12,76.97) and (385.51,121.36) .. (421.9,115.6) ;
%Shape: Ellipse [id:dp7681677659184208] 
\draw   (413.3,126.29) .. controls (413.3,120.39) and (417.15,115.6) .. (421.9,115.6) .. controls (426.65,115.6) and (430.5,120.39) .. (430.5,126.29) .. controls (430.5,132.19) and (426.65,136.98) .. (421.9,136.98) .. controls (417.15,136.98) and (413.3,132.19) .. (413.3,126.29) -- cycle ;
%Curve Lines [id:da13925019658887294] 
\draw    (92.45,124.17) .. controls (114.99,115.95) and (174.05,161.76) .. (41.94,176.02) ;

% Text Node
\draw (211.18,133.42) node [anchor=north west][inner sep=0.75pt]  [xscale=1,yscale=1]  {$\cong$};
% Text Node
\draw (8.35,83.29) node [anchor=north west][inner sep=0.75pt]  [xscale=1,yscale=1]  {$\sigma^i$};
% Text Node
\draw (8.35,129.96) node [anchor=north west][inner sep=0.75pt]  [xscale=1,yscale=1]  {$\sigma^j$};
% Text Node
\draw (8.35,183.55) node [anchor=north west][inner sep=0.75pt]   [xscale=1,yscale=1]  {$\sigma^k$};
% Text Node
\draw (252.37,84.16) node [anchor=north west][inner sep=0.75pt]   [xscale=1,yscale=1]  {$\sigma^i$};
% Text Node
\draw (253.14,135.15) node [anchor=north west][inner sep=0.75pt]   [xscale=1,yscale=1]  {$\sigma^j$};
% Text Node
\draw (254.7,180.95) node [anchor=north west][inner sep=0.75pt]   [xscale=1,yscale=1]  {$\sigma^k$};
% Text Node
\draw (106.41,239.73) node [anchor=north west][inner sep=0.75pt]   [xscale=1,yscale=1]  {$\left(\sigma^i\cdot \sigma^j\right)\cdot \sigma^k=\sigma^i\cdot\left( \sigma^j\cdot \sigma^k \right)$};

\end{tikzpicture}
\end{center}
 and the commutativity 
 \begin{center}
\tikzset{every picture/.style={line width=0.75pt}} %set default line width to 0.75pt        

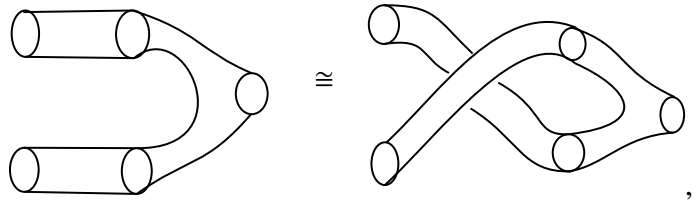
\begin{tikzpicture}[x=0.75pt,y=0.75pt,yscale=-1,xscale=1]

%Flowchart: Connector [id:dp2962469666354113] 
\draw   (57,131.67) .. controls (57,125.03) and (60.74,119.65) .. (65.35,119.65) .. controls (69.96,119.65) and (73.7,125.03) .. (73.7,131.67) .. controls (73.7,138.31) and (69.96,143.69) .. (65.35,143.69) .. controls (60.74,143.69) and (57,138.31) .. (57,131.67) -- cycle ;
%Flowchart: Connector [id:dp36928231238916487] 
\draw   (59.91,200.69) .. controls (59.91,194.31) and (63.29,189.14) .. (67.46,189.14) .. controls (71.63,189.14) and (75.01,194.31) .. (75.01,200.69) .. controls (75.01,207.07) and (71.63,212.24) .. (67.46,212.24) .. controls (63.29,212.24) and (59.91,207.07) .. (59.91,200.69) -- cycle ;
%Flowchart: Connector [id:dp46425367253310545] 
\draw   (117.41,161.65) .. controls (117.41,155.93) and (121,151.3) .. (125.44,151.3) .. controls (129.87,151.3) and (133.46,155.93) .. (133.46,161.65) .. controls (133.46,167.37) and (129.87,172) .. (125.44,172) .. controls (121,172) and (117.41,167.37) .. (117.41,161.65) -- cycle ;
%Curve Lines [id:da9194522781918303] 
\draw    (65.35,119.65) .. controls (98.52,130.08) and (99,140.12) .. (125.44,151.3) ;
%Curve Lines [id:da367577985480792] 
\draw    (67.46,212.24) .. controls (93.4,193.61) and (101.94,198.83) .. (125.09,172) ;
%Curve Lines [id:da48758212225173525] 
\draw    (66.15,143.69) .. controls (92.42,121.33) and (122.3,189.89) .. (67.46,189.14) ;
%Flowchart: Connector [id:dp8952703070235815] 
\draw   (184.48,126.77) .. controls (184.48,121.37) and (187.89,117) .. (192.1,117) .. controls (196.31,117) and (199.72,121.37) .. (199.72,126.77) .. controls (199.72,132.16) and (196.31,136.54) .. (192.1,136.54) .. controls (187.89,136.54) and (184.48,132.16) .. (184.48,126.77) -- cycle ;
%Flowchart: Connector [id:dp4517173812105889] 
\draw   (185.81,196.98) .. controls (185.81,191.02) and (188.85,186.19) .. (192.6,186.19) .. controls (196.35,186.19) and (199.39,191.02) .. (199.39,196.98) .. controls (199.39,202.93) and (196.35,207.76) .. (192.6,207.76) .. controls (188.85,207.76) and (185.81,202.93) .. (185.81,196.98) -- cycle ;
%Flowchart: Connector [id:dp23844262007599504] 
\draw   (280.72,136.54) .. controls (280.72,132.04) and (283.65,128.4) .. (287.26,128.4) .. controls (290.88,128.4) and (293.81,132.04) .. (293.81,136.54) .. controls (293.81,141.03) and (290.88,144.68) .. (287.26,144.68) .. controls (283.65,144.68) and (280.72,141.03) .. (280.72,136.54) -- cycle ;
%Curve Lines [id:da05596793358648611] 
\draw    (193.43,186.19) .. controls (219.93,161.77) and (251.9,112.93) .. (289.17,128.4) ;
%Curve Lines [id:da38961955093082723] 
\draw    (192.6,207.76) .. controls (219.1,183.34) and (262.5,124.33) .. (288.51,146.31) ;
%Curve Lines [id:da14534185451788162] 
\draw    (192.1,117) .. controls (218.27,117) and (218.27,123.51) .. (236.83,141.42) ;
%Curve Lines [id:da7278422189390561] 
\draw    (192.1,136.54) .. controls (214.96,133.28) and (209.66,143.05) .. (224.9,151.19) ;
%Curve Lines [id:da44367883226630944] 
\draw    (235.83,169.1) .. controls (260.68,184.56) and (261.34,201.66) .. (286.52,200.85) ;
%Curve Lines [id:da9694405330399225] 
\draw    (249.75,156.07) .. controls (274.59,169.91) and (267.3,182.94) .. (285.19,182.12) ;
%Flowchart: Connector [id:dp269728259214428] 
\draw   (277.24,191.48) .. controls (277.24,186.31) and (280.8,182.12) .. (285.19,182.12) .. controls (289.58,182.12) and (293.14,186.31) .. (293.14,191.48) .. controls (293.14,196.65) and (289.58,200.85) .. (285.19,200.85) .. controls (280.8,200.85) and (277.24,196.65) .. (277.24,191.48) -- cycle ;
%Curve Lines [id:da1334292545269744] 
\draw    (287.26,128.4) .. controls (314.35,137.35) and (308.98,149.59) .. (338.86,163.4) ;
%Curve Lines [id:da3324944048747943] 
\draw    (285.19,182.12) .. controls (328.26,177.24) and (315.67,158.52) .. (288.51,146.31) ;
%Curve Lines [id:da007334003085733509] 
\draw    (285.19,200.85) .. controls (302.94,198.35) and (312.43,182.95) .. (338.2,181.31) ;
%Flowchart: Connector [id:dp4921408097186748] 
\draw   (331.57,172.35) .. controls (331.57,167.41) and (334.24,163.4) .. (337.54,163.4) .. controls (340.83,163.4) and (343.5,167.41) .. (343.5,172.35) .. controls (343.5,177.3) and (340.83,181.31) .. (337.54,181.31) .. controls (334.24,181.31) and (331.57,177.3) .. (331.57,172.35) -- cycle ;
%Flowchart: Connector [id:dp9284680094187208] 
\draw   (4.36,131.35) .. controls (4.36,124.89) and (7.46,119.65) .. (11.27,119.65) .. controls (15.08,119.65) and (18.18,124.89) .. (18.18,131.35) .. controls (18.18,137.81) and (15.08,143.05) .. (11.27,143.05) .. controls (7.46,143.05) and (4.36,137.81) .. (4.36,131.35) -- cycle ;
%Flowchart: Connector [id:dp7352239199345547] 
\draw   (3.5,200.06) .. controls (3.5,193.92) and (6.74,188.94) .. (10.74,188.94) .. controls (14.74,188.94) and (17.98,193.92) .. (17.98,200.06) .. controls (17.98,206.2) and (14.74,211.18) .. (10.74,211.18) .. controls (6.74,211.18) and (3.5,206.2) .. (3.5,200.06) -- cycle ;
%Straight Lines [id:da16498920942198103] 
\draw    (10.26,120.46) -- (65.35,119.65) ;
%Straight Lines [id:da7682904954313795] 
\draw    (11.05,143.05) -- (66.15,143.69) ;
%Straight Lines [id:da42024074986208804] 
\draw    (10.74,188.94) -- (67.46,189.14) ;
%Straight Lines [id:da32816006648383167] 
\draw    (10.74,211.18) -- (67.46,212.24) ;

% Text Node
\draw (154.96,149.29) node [anchor=north west][inner sep=0.75pt]  [xscale=1,yscale=1]  {$\cong$};

\end{tikzpicture},
\end{center}
one can observe that the algebraic structure of ${\cal C}$, forms a commutative Frobenius algebra, as mentioned earlier.

It is now sufficient to compute arbitrary transition amplitudes using the Frobenius algebra, since any oriented Riemann surface can be constructed from the elementary bordisms listed above. However, the same surface admits multiple distinct decompositions. These different decompositions could introduce additional algebraic relations among the generators $\textit{m}$, ${\Delta}$, $\rm{I}$, and $\tau$, but the `sewing theorem' ensures that no further relations arise. For a more detailed discussion, see \cite{Moore:2015TQFT}.  

\noindent{\textbf{Genus-$g$ partition function}}: A convenient approach to computing the genus-$g$ partition function is to introduce a `handle-adding operator' ${\cal H}$ within the bordism formalism:
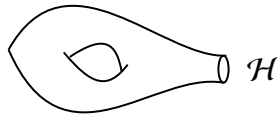
\begin{center}
\tikzset{every picture/.style={line width=0.75pt}} %set default line width to 0.75pt        

\begin{tikzpicture}[x=0.75pt,y=0.75pt,yscale=-1,xscale=1]

%Curve Lines [id:da21827534496805567] 
\draw    (35.38,191.81) .. controls (52.8,180.15) and (58.9,184.43) .. (61.51,199.18) ;
%Curve Lines [id:da23986022179345512] 
\draw    (32.33,189.48) .. controls (46.27,206.18) and (54.54,208.89) .. (64.56,195.69) ;
%Curve Lines [id:da9197804107036163] 
\draw    (4.5,188.15) .. controls (33.8,135.9) and (82.13,192.69) .. (114.36,191.17) ;
%Curve Lines [id:da8393768115239479] 
\draw    (4.5,188.15) .. controls (32.33,251.75) and (77.74,198.75) .. (112.89,204.8) ;
%Flowchart: Connector [id:dp7009519665542019] 
\draw   (110.33,197.99) .. controls (110.33,194.23) and (111.48,191.17) .. (112.89,191.17) .. controls (114.31,191.17) and (115.46,194.23) .. (115.46,197.99) .. controls (115.46,201.75) and (114.31,204.8) .. (112.89,204.8) .. controls (111.48,204.8) and (110.33,201.75) .. (110.33,197.99) -- cycle ;
% Text Node
\draw (123.37,190.39) node [anchor=north west][inner sep=0.75pt]  [xscale=1,yscale=1]  {${\cal H}$ };
\end{tikzpicture}.
\end{center}
This bordism is a composition $\textit{m}\circ \Delta$:
\begin{equation}\label{Hdl:Funt}
1\rightarrow \sum_{ij} \Delta_{ij}\sigma^i\otimes\sigma^j\rightarrow \sum_{ij}\Delta_{ij}\sigma^{i} \cdot \sigma^j,
\end{equation}
yielding ${\cal H}=\sum_{ij}\Delta_{ij}\sigma^{i} \cdot \sigma^j$. The amplitude for the genus-$g$ surface would be given by 
\begin{equation}\label{TA:Funt}
Z(\Sigma_g)=\tau({\cal H}^g).
\end{equation}
For a \textit{semi-simple} Frobenius algebra, there exists a basis of commuting idempotents $\{e_1,\ldots, e_n\}\in {\cal C}$, that is $e_ie_j=\delta_{ij}e_j$. From the torus partition function $Z(T^2)=n$, we deduce that the handle-adding operator can be expressed as:
\begin{equation}\label{Hdl2:Funt}
{\cal H}=\sum_{i}\Delta_{ii}e_i=\sum_i \frac{1}{\lambda_i}e_i,
\end{equation}
where $\lambda_i=\tau(e_i)\neq 0$. Therefore, we obtain the genus-$g$ partition function:
\begin{equation}\label{TA2:Funt}
Z(\Sigma_g)=\sum_i\lambda_i^{1-g}.
\end{equation}

\subsection{Comments on Hilbert spaces in TQFTs}\label{HS:TQFT}

The complex vector space defined above is equipped with a bilinear inner product. However, it is not sufficient to claim that it is a Hilbert space, since the inner product of the vector space is, in general, not positive-definite. Crucially, unlike in a Hilbert space, this inner product is bilinear rather than sesquilinear. It is often assumed to be a Hilbert space because it corresponds to the zero-energy sector of a physical quantum field theory. However, this assumption is oversimplified. The subtlety lies in the method of projecting the full physical state space onto the ground states. For example, a 2D Witten-type TQFT may lack a positive-definite inner product. To illustrate this, consider a topological sigma model with target space $\mathbb{CP} ^1$, whose quantum cohomology provides the vector space which has two elements: $\{1, \sigma\}$, where $\sigma$ represents the hyperplane class of $\mathbb{CP} ^1$. The standard inner product defined via integration of the top-form over $\mathbb{CP} ^1$ yields the matrix 
\begin{equation*}
\begin{pmatrix}
0 & 1 \\
1 & 0 \\
\end{pmatrix},
\end{equation*}
which is not positive-definite. This demonstrates that even in well-behaved TQFTs, the zero-energy sector need not inherit a Hilbert space structure from the full theory. One might argue that Witten-type TQFTs are non-unitary as quantum theories. However, this conclusion is premature because it does not necessarily imply that the associated sesquilinear form fails to be positive definite. A detailed analysis of this point will be presented in \ref{BWTQFT}. For a complex Hilbert space in this context, the unitary structure (e.g., the compatibility of the inner product with bordism operations) must be carefully constructed. We will review this construction in the next subsection.

\subsection{Unitarity}
As emphasized at the beginning of Section \ref{AS:axioms}, the algebraic structure of the vector space (including its inner product) is fully determined by the topology of bordisms. To qualify as a unitary theory in the standard quantum mechanical sense, the vector space must form a complex Hilbert space. Furthermore, the TQFT's functorial assignment $Z(\Sigma)$ must be a unitary operator for any bordism $\Sigma$. 

We begin by analyzing particle-antiparticle annihilation and pair production in this framework. If a circle $S^1$ represents a particle, then it is natural to expect that its reversed orientation $(S^1)^\vee$ is an anti-particle. The particle and anti-particle annihilation is topologically encoded as bordism $S^1 \coprod (S^1)^\vee \rightarrow \emptyset$, where $\emptyset$ is the empty 1-dimensional manifold. This defines a quadratic form
\begin{equation}\label{QF:Funt}
Q_{S^1}: {\cal C}(S^1)\otimes{\cal C}((S^1)^\vee)\rightarrow \mathbb{ C} .
\end{equation}
For the theory to be physical, $Q_{S^1}$ must be positive-definite, ensuring unitarity in quantum amplitudes. Similarly, pair production is represented by reversed bordism $\emptyset \rightarrow S^1 \coprod (S^1)^\vee$, inducing a map  \begin{equation}\label{PD:Funt}
\Delta_{S^1}: \mathbb{ C} \rightarrow {\cal C}(S^1)\otimes{\cal C}((S^1)^\vee) .
\end{equation}
By analyzing the S-diagram formed by composing $Q_{S^1}$ and $\Delta_{S^1}$,
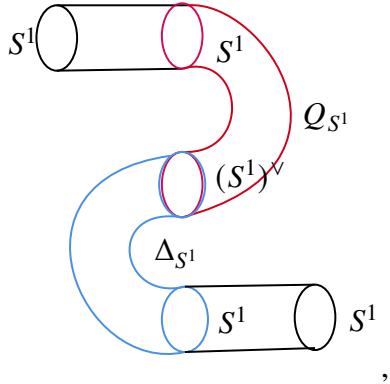
\begin{center}

\tikzset{every picture/.style={line width=0.75pt}} %set default line width to 0.75pt        

\begin{tikzpicture}[x=0.75pt,y=0.75pt,yscale=-1,xscale=1]

%Flowchart: Connector [id:dp258016056469313] 
\draw   (21,98.5) .. controls (21,89.39) and (25.59,82) .. (31.25,82) .. controls (36.91,82) and (41.5,89.39) .. (41.5,98.5) .. controls (41.5,107.61) and (36.91,115) .. (31.25,115) .. controls (25.59,115) and (21,107.61) .. (21,98.5) -- cycle ;
%Straight Lines [id:da15441882298907328] 
\draw    (31.25,82) -- (97.5,82) ;
%Straight Lines [id:da017336044662630723] 
\draw    (31.25,115) -- (94.5,114) ;
%Flowchart: Connector [id:dp8015547560002184] 
\draw  [color={rgb, 255:red, 208; green, 2; blue, 81 }  ,draw opacity=0.96 ] (84.25,97.5) .. controls (84.25,88.39) and (88.84,81) .. (94.5,81) .. controls (100.16,81) and (104.75,88.39) .. (104.75,97.5) .. controls (104.75,106.61) and (100.16,114) .. (94.5,114) .. controls (88.84,114) and (84.25,106.61) .. (84.25,97.5) -- cycle ;
%Curve Lines [id:da8065310077805049] 
\draw [color={rgb, 255:red, 208; green, 2; blue, 27 }  ,draw opacity=1 ]   (97.5,82) .. controls (135.5,79) and (193.5,163) .. (94.5,189) ;
%Flowchart: Connector [id:dp10791787789356133] 
\draw  [color={rgb, 255:red, 208; green, 2; blue, 81 }  ,draw opacity=0.96 ] (84.25,172.5) .. controls (84.25,163.39) and (88.84,156) .. (94.5,156) .. controls (100.16,156) and (104.75,163.39) .. (104.75,172.5) .. controls (104.75,181.61) and (100.16,189) .. (94.5,189) .. controls (88.84,189) and (84.25,181.61) .. (84.25,172.5) -- cycle ;
%Flowchart: Connector [id:dp342780451082209] 
\draw  [color={rgb, 255:red, 74; green, 144; blue, 226 }  ,draw opacity=1 ] (84.25,240.5) .. controls (84.25,231.39) and (89.45,224) .. (95.88,224) .. controls (102.3,224) and (107.5,231.39) .. (107.5,240.5) .. controls (107.5,249.61) and (102.3,257) .. (95.88,257) .. controls (89.45,257) and (84.25,249.61) .. (84.25,240.5) -- cycle ;
%Curve Lines [id:da21943889518004656] 
\draw [color={rgb, 255:red, 208; green, 2; blue, 27 }  ,draw opacity=1 ]   (94.5,114) .. controls (125.5,105) and (130.5,158) .. (94.5,156) ;
%Flowchart: Connector [id:dp7052660702811009] 
\draw  [color={rgb, 255:red, 74; green, 144; blue, 226 }  ,draw opacity=1 ] (82.88,172.5) .. controls (82.88,163.39) and (88.08,156) .. (94.5,156) .. controls (100.92,156) and (106.13,163.39) .. (106.13,172.5) .. controls (106.13,181.61) and (100.92,189) .. (94.5,189) .. controls (88.08,189) and (82.88,181.61) .. (82.88,172.5) -- cycle ;
%Curve Lines [id:da6300454351286042] 
\draw [color={rgb, 255:red, 74; green, 144; blue, 226 }  ,draw opacity=1 ]   (88.5,158) .. controls (0.5,170) and (43.5,269) .. (95.88,257) ;
%Curve Lines [id:da8664358571407347] 
\draw [color={rgb, 255:red, 74; green, 144; blue, 226 }  ,draw opacity=1 ]   (94.5,189) .. controls (55.5,183) and (62.5,232) .. (95.88,224) ;
%Straight Lines [id:da004145573113350731] 
\draw    (95.88,224) -- (161.5,223) ;
%Straight Lines [id:da7609792107789711] 
\draw    (95.88,257) -- (161.5,255) ;
%Flowchart: Connector [id:dp36977955560464126] 
\draw   (151.25,239.5) .. controls (151.25,230.39) and (155.84,223) .. (161.5,223) .. controls (167.16,223) and (171.75,230.39) .. (171.75,239.5) .. controls (171.75,248.61) and (167.16,256) .. (161.5,256) .. controls (155.84,256) and (151.25,248.61) .. (151.25,239.5) -- cycle ;

% Text Node
\draw (5,91.4) node [anchor=north west][inner sep=0.75pt]  [xscale=1,yscale=1]  {$S^1$};
% Text Node
\draw (110,95.4) node [anchor=north west][inner sep=0.75pt]  [xscale=1,yscale=1]  {$S^1$};
% Text Node
\draw (154,129.4) node [anchor=north west][inner sep=0.75pt]  [xscale=1,yscale=1]  {$Q_{S^1}$};
% Text Node
\draw (109,158.4) node [anchor=north west][inner sep=0.75pt]  [xscale=1,yscale=1]  {$(S^1)^{\vee}$};
% Text Node
\draw (79,198.4) node [anchor=north west][inner sep=0.75pt]  [xscale=1,yscale=1]  {$\Delta_{S^1}$};
% Text Node
\draw (111,232.4) node [anchor=north west][inner sep=0.75pt]  [xscale=1,yscale=1]  {$S^1$};
% Text Node
\draw (177,231.4) node [anchor=north west][inner sep=0.75pt]  [xscale=1,yscale=1]  {$S^1$};

\end{tikzpicture}, 
\end{center}
one can show that the quadratic form (\ref{QF:Funt}) is also non-degenerate. This establishes an isomorphism ${\cal C}((S^1)^\vee)\cong{\cal C}(S^1)^\vee$, confirming the perfect duality between particles and anti-particles. 

Given that ${\cal C}$ is a Hilbert space, its sesquilinear form induces an antilinear map ${\cal C}^\vee\rightarrow \bar{{\cal C}}$. Therefore, in a unitary TQFT, the orientation reversal of $S^1$ corresponds to the conjugation of the Hilbert space complex 
\begin{equation}\label{OC}
{\cal C}((S^1)^\vee)\cong \bar{\cal C}(S^1).
\end{equation}
Recall $Z(\Sigma)$ in \ref{Tran:Funt}, given a bordism $\Sigma$, its orientation reversal followed by dualization yields a new bordism
\begin{equation}\nonumber
\bar{\Sigma}^\vee: {\cal C}^{\otimes q}\xrightarrow \quad{\cal C}^{\otimes p}.
\end{equation}
The additional condition required for unitary theories is the compatibility condition:
\begin{equation}\label{OC2}
Z(\bar{\Sigma}^\vee)= Z(\Sigma)^{\dagger},
\end{equation}
where $\dagger$ denotes the Hermitian conjugation. This adjoint operation physically corresponds to the time-reversal transformation that exchanges initial and final states in the quantum evolution.

Finally, the torus partition function is constructed through $Q_{S^1}\circ\Delta_{S^1}$.  We call a unitary TQFT `real' when its partition functions are positive real-valued (i.e., $\lambda_i>0$ in Eq.(\ref{TA2:Funt}). The authors of \cite{Durhuus:1993cq} argue that (\ref{TA2:Funt}) classifies such 2D TQFTs.

\noindent{\textbf{The Verlinde algebra of $U(1)_{n+1}$}: This is a unitary theory with the vacuum equation as follows:
\begin{equation}\label{VEVA}
e^{i(n+1)\varphi}=1.
\end{equation}
The idempotent basis elements are:
\begin{equation}\label{VAIB}
e_k=\frac{1}{n+1}\sum^{n}_{m=0} e^{im\left(\varphi-\frac{2\pi (k-1)}{n+1}\right)}, \quad k=1,\ldots,n+1.
\end{equation}
The ground states are given by
\begin{equation}\label{VAGS}
\ket{e_1},\ldots, \ket{e_{n+1}}.
\end{equation}
with their dual states:
\begin{equation}\label{DVAGS}
\bra{\bar{e}_1},\ldots, \bra{\bar{e}_{n+1}}.
\end{equation}
The trace condition produces identical eigenvalues:
\begin{equation}\nonumber
\lambda_i=\tau(e_i)=\frac{1}{n+1}.
\end{equation}
The genus-$g$ partition function is:
\begin{equation}\nonumber
Z(\Sigma_g)=(n+1)^g.
\end{equation}

\section{Bordism categories in Witten-type 2D TQFTs}\label{BWTQFT}
The bordisms analyzed in the previous section are incompatible with Witten-type TQFTs due to the fundamental constraints in this framework. First, their topological structure fails to satisfy the necessary conditions for functoriality. For
example, the following two bordisms  
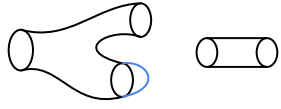
\begin{center}

\tikzset{every picture/.style={line width=0.75pt}} %set default line width to 0.75pt        

\begin{tikzpicture}[x=0.75pt,y=0.75pt,yscale=-1,xscale=1]

%Flowchart: Connector [id:dp6052557676542344] 
\draw   (9,213.61) .. controls (9,207.93) and (11.75,203.32) .. (15.14,203.32) .. controls (18.52,203.32) and (21.27,207.93) .. (21.27,213.61) .. controls (21.27,219.3) and (18.52,223.9) .. (15.14,223.9) .. controls (11.75,223.9) and (9,219.3) .. (9,213.61) -- cycle ;
%Curve Lines [id:da026165632711500963] 
\draw    (15.14,203.32) .. controls (39.55,209.98) and (53.64,190.61) .. (75.58,190) ;
%Flowchart: Connector [id:dp8046164597905234] 
\draw   (70.29,198.48) .. controls (70.29,193.79) and (72.66,190) .. (75.58,190) .. controls (78.49,190) and (80.86,193.79) .. (80.86,198.48) .. controls (80.86,203.16) and (78.49,206.95) .. (75.58,206.95) .. controls (72.66,206.95) and (70.29,203.16) .. (70.29,198.48) -- cycle ;
%Curve Lines [id:da25977476239043384] 
\draw    (15.14,223.9) .. controls (34.32,218.46) and (41.11,243.88) .. (66.18,237.22) ;
%Flowchart: Connector [id:dp9077667347763828] 
\draw   (60.69,228.75) .. controls (60.69,224.07) and (63.15,220.27) .. (66.18,220.27) .. controls (69.2,220.27) and (71.66,224.07) .. (71.66,228.75) .. controls (71.66,233.43) and (69.2,237.22) .. (66.18,237.22) .. controls (63.15,237.22) and (60.69,233.43) .. (60.69,228.75) -- cycle ;
%Curve Lines [id:da731422000278415] 
\draw    (75.58,206.95) .. controls (60.43,201.5) and (39.02,213.61) .. (66.18,220.27) ;
%Curve Lines [id:da3853815656200781] 
\draw [color={rgb, 255:red, 74; green, 131; blue, 226 }  ,draw opacity=1 ]   (66.18,220.27) .. controls (79.75,218.46) and (87.58,232.99) .. (66.18,237.22) ;
%Flowchart: Connector [id:dp5652214306102366] 
\draw   (103.77,214.82) .. controls (103.77,210.81) and (106.11,207.56) .. (108.99,207.56) .. controls (111.88,207.56) and (114.21,210.81) .. (114.21,214.82) .. controls (114.21,218.84) and (111.88,222.09) .. (108.99,222.09) .. controls (106.11,222.09) and (103.77,218.84) .. (103.77,214.82) -- cycle ;
%Flowchart: Connector [id:dp28038028147050476] 
\draw   (134.06,214.82) .. controls (134.06,210.81) and (136.39,207.56) .. (139.28,207.56) .. controls (142.16,207.56) and (144.5,210.81) .. (144.5,214.82) .. controls (144.5,218.84) and (142.16,222.09) .. (139.28,222.09) .. controls (136.39,222.09) and (134.06,218.84) .. (134.06,214.82) -- cycle ;
%Straight Lines [id:da19195978090409593] 
\draw    (108.99,207.56) -- (139.28,207.56) ;
%Straight Lines [id:da6956294130482391] 
\draw    (108.99,222.09) -- (139.28,222.09) ;

\end{tikzpicture}

\end{center}
shall both represent the identity map $\cal C\rightarrow\cal C$. However, the left diagram fails to satisfy this condition, since in generic Witten-type TQFTs it corresponds to the map ${{\cal C}}\rightarrow {{\cal C}}\otimes \tau(1)={{\cal C}}\otimes 0=0 $.  This inconsistency precludes their use in Witten-type constructions.  Secondly, as discussed in Section \ref{AS:axioms}, these bordisms would suggest that Witten-type TQFTs ``were non-unitary" in the following more refined discussion. Consider the GLSM for $\mathbb{CP}^{n}$, whose vacuum equation is given by:
\begin{equation}\nonumber
\sigma^{n+1}=\Lambda^{n+1},
\end{equation}
where $\sigma$ is a complex scalar and $\Lambda$ is the complex-valued dynamic scale. In the UV regime, this relation can be expressed as $\sigma^{n+1}=\Lambda^{n+1}=q=e^{-t}$, where $q$ is the Novikov variable and $t$ is the complexified FI-parameter. The twisted chiral ring is generated by the basis elements
\begin{equation}\nonumber
1, \sigma\Lambda^{-1}, \ldots, \sigma^{n}\Lambda^{-n}.
\end{equation}
Through the operator-state correspondence, the ground states are given by:
\begin{equation}\nonumber
\ket{0},\ldots, \ket{k}=\sigma^k\Lambda^{-k}\ket{0}, \ldots, \ket{n}
\end{equation}
The vacuum equation also admits an anti-holomorphic representation:
\begin{equation}\nonumber
\bar{\sigma}^{n+1}=\bar{\Lambda}^{n+1}.
\end{equation}
Similarly, the anti-twisted chiral ring is generated by:
\begin{equation}\nonumber
1, \bar{\sigma}\bar{\Lambda}^{-1}, \ldots, \bar{\sigma}^{n}\bar{\Lambda}^{-n}.
\end{equation}
The corresponding ground states are given by:
\begin{equation}\nonumber
\ket{0},\ldots, \ket{\bar{k}}=\bar{\sigma}^k\bar{\Lambda}^{-k}\ket{\bar{1}}, \ldots, \ket{\bar{n}}.
\end{equation}
In the large-$\Lambda$ limit ($|\Lambda|\gg 1$), the fluctuations of the domain walls are suppressed \cite{Witten:1978bc,Gu:2022ugf}. This implies that each vacuum state effectively corresponds to that of a free massive theory, yielding \cite{Cecotti:1991vb}
\begin{equation}\nonumber
\bra{\bar{i}}\ket{j}=A\delta_{\bar{i}j},
\end{equation}
where $A=\bra{0}\ket{0}$. On the basis of the above analysis, one can identify the following operator relation:
\begin{equation}\nonumber
\bar{\sigma}^{k}\bar{\Lambda}^{-k}=\sigma^{n+1-k}\Lambda^{-(n+1-k)},
\end{equation}
which is directly derivable from the vacuum equations. When considering the topological A-twisted theory, we may formally define the conjugate state as follows\footnote{The sesquilinear structure can be seen as the following: If we consider an operator like $c\cdot\sigma^{k}\Lambda^{-k}$, then we expect that the following operator $\bar{c}\cdot\sigma^{n+1-k}\Lambda^{-(n+1-k)}$ would annihilate it to give a real constant operator.}:
\begin{equation}\nonumber 
\ket{\bar{i}}\coloneqq\ket{n+1-i}.
\end{equation}
However, we have the inner product relation in the A-twisted theory:
\begin{equation}\nonumber
\bra{n-j}\ket{j}=\frac{1}{\Lambda^n},
\end{equation}
which implies the following metric structure:
\begin{equation}\nonumber
\bra{\bar{i}}\ket{j}=\frac{1}{\Lambda^n}\delta_{i,j+1}, \quad i,j\in \mathbb{Z}/(n+1)\mathbb{Z}
\end{equation}
Since this metric fails to satisfy the hermiticity condition, the theory looks potentially non-unitary.

\noindent{\textbf{Example}}:  The metric for $\mathbb{CP}^2$  takes the form:
\begin{equation}\nonumber
g_{\bar{i}j}=\begin{pmatrix}
 &  & \frac{1}{\Lambda^2} \\
 \frac{1}{\Lambda^2}&  & \\
 &  \frac{1}{\Lambda^2}&  \\
\end{pmatrix}.
\end{equation}
While readers might accept this at face value, we must recall the correspondence mentioned in Section \ref{Intr}: \textit{Quantum cohomology of $\mathbb{CP}^n$ $\cong$ Verlinde algebra of $U(1)_{n+1}$}. This presents a conceptual difficulty, as we cannot naturally expect a unitary quantum field theory to correspond to a manifestly non-unitary theory through this correspondence.

One potential resolution to these issues involves redefining both the inner product structure and the associated bordism operations. We now introduce the following fundamental bordisms for Witten-type TQFTs:
\begin{itemize}
    \item The multiplication and co-multiplication bordisms remain unchanged from our previous definition, which we restate here for completeness:

    \begin{center}

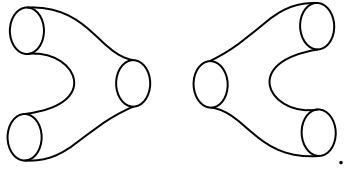
\tikzset{every picture/.style={line width=0.75pt}} %set default line width to 0.75pt        

\begin{tikzpicture}[x=0.75pt,y=0.75pt,yscale=-1,xscale=1]

%Flowchart: Connector [id:dp10863301384474677] 
\draw   (4.16,189.55) .. controls (4.16,183.06) and (8.01,177.81) .. (12.76,177.81) .. controls (17.5,177.81) and (21.35,183.06) .. (21.35,189.55) .. controls (21.35,196.03) and (17.5,201.29) .. (12.76,201.29) .. controls (8.01,201.29) and (4.16,196.03) .. (4.16,189.55) -- cycle ;
%Flowchart: Connector [id:dp7800812273951812] 
\draw   (3,243.3) .. controls (3,236.82) and (6.85,231.56) .. (11.59,231.56) .. controls (16.34,231.56) and (20.18,236.82) .. (20.18,243.3) .. controls (20.18,249.78) and (16.34,255.04) .. (11.59,255.04) .. controls (6.85,255.04) and (3,249.78) .. (3,243.3) -- cycle ;
%Flowchart: Connector [id:dp14773860375590286] 
\draw   (57.75,216.11) .. controls (57.75,209.63) and (61.59,204.38) .. (66.34,204.38) .. controls (71.08,204.38) and (74.93,209.63) .. (74.93,216.11) .. controls (74.93,222.6) and (71.08,227.85) .. (66.34,227.85) .. controls (61.59,227.85) and (57.75,222.6) .. (57.75,216.11) -- cycle ;
%Curve Lines [id:da6232442504926348] 
\draw    (12.76,177.81) .. controls (44.06,176.57) and (48.72,201.29) .. (66.34,204.38) ;
%Curve Lines [id:da9156126975008196] 
\draw    (11.59,255.04) .. controls (34.74,256.28) and (39.4,239.59) .. (66.34,227.85) ;
%Curve Lines [id:da42041247753338706] 
\draw    (12.76,201.29) .. controls (36.49,198.2) and (52.22,226.62) .. (11.59,231.56) ;
%Flowchart: Connector [id:dp9498884933536151] 
\draw   (168.49,240.65) .. controls (168.72,247.13) and (165.07,252.54) .. (160.33,252.73) .. controls (155.58,252.92) and (151.55,247.82) .. (151.32,241.35) .. controls (151.09,234.87) and (154.74,229.46) .. (159.48,229.27) .. controls (164.22,229.08) and (168.26,234.17) .. (168.49,240.65) -- cycle ;
%Flowchart: Connector [id:dp5848691350590072] 
\draw   (167.72,186.89) .. controls (167.95,193.37) and (164.3,198.77) .. (159.55,198.97) .. controls (154.81,199.16) and (150.78,194.06) .. (150.55,187.58) .. controls (150.31,181.11) and (153.97,175.7) .. (158.71,175.51) .. controls (163.45,175.31) and (167.48,180.41) .. (167.72,186.89) -- cycle ;
%Flowchart: Connector [id:dp6344263278805277] 
\draw   (113.99,216.27) .. controls (114.22,222.75) and (110.57,228.16) .. (105.83,228.35) .. controls (101.08,228.54) and (97.05,223.45) .. (96.82,216.97) .. controls (96.58,210.49) and (100.24,205.08) .. (104.98,204.89) .. controls (109.72,204.7) and (113.76,209.8) .. (113.99,216.27) -- cycle ;
%Curve Lines [id:da33557133479942003] 
\draw    (160.33,252.73) .. controls (129.09,255.23) and (123.54,230.73) .. (105.83,228.35) ;
%Curve Lines [id:da6925555797259483] 
\draw    (158.71,175.51) .. controls (135.53,175.21) and (131.47,192.07) .. (104.98,204.89) ;
%Curve Lines [id:da9730843444107083] 
\draw    (159.48,229.27) .. controls (135.88,233.32) and (119.14,205.55) .. (159.55,198.97) ;

\end{tikzpicture}.
\end{center}
\item The unit and co-unit maps are given by

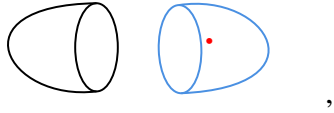
\begin{center}

\tikzset{every picture/.style={line width=0.75pt}} %set default line width to 0.75pt        

\begin{tikzpicture}[x=0.75pt,y=0.75pt,yscale=-1,xscale=1]

%Curve Lines [id:da42227886447743124] 
\draw    (48.75,165.5) .. controls (-23.5,166.5) and (3.5,215.5) .. (48.75,210) ;
%Flowchart: Connector [id:dp833556052694024] 
\draw   (38,187.75) .. controls (38,175.46) and (42.81,165.5) .. (48.75,165.5) .. controls (54.69,165.5) and (59.5,175.46) .. (59.5,187.75) .. controls (59.5,200.04) and (54.69,210) .. (48.75,210) .. controls (42.81,210) and (38,200.04) .. (38,187.75) -- cycle ;
%Curve Lines [id:da11969617127657828] 
\draw [color={rgb, 255:red, 74; green, 144; blue, 226 }  ,draw opacity=1 ]   (91.54,210.92) .. controls (163.73,207.79) and (135.3,159.61) .. (90.23,166.44) ;
%Flowchart: Connector [id:dp542938163927856] 
\draw  [color={rgb, 255:red, 74; green, 144; blue, 226 }  ,draw opacity=1 ] (101.63,188.36) .. controls (101.99,200.65) and (97.47,210.75) .. (91.54,210.92) .. controls (85.61,211.1) and (80.5,201.28) .. (80.14,189) .. controls (79.78,176.71) and (84.29,166.61) .. (90.23,166.44) .. controls (96.16,166.26) and (101.27,176.08) .. (101.63,188.36) -- cycle ;

% Text Node
\draw (103,182.4) node [anchor=north west][inner sep=0.75pt]  [xscale=0.5,yscale=0.5]  {$\color{red} \bullet$};

\end{tikzpicture},

\end{center}
respectively.

\end{itemize}
In comparison with the bordisms defined in Section \ref{AS:axioms}, we observe that only the co-unit map incorporates an additional operator insertion (denoted by $\color{red}\bullet$), which we term the ``background particle" throughout this work. The bordisms in Section \ref{AS:axioms} correspond to the special case where this insertion reduces to the identity operator. The background particle must satisfy the non-degeneracy condition expressed through the two-sphere ``partition function:"
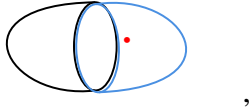
\begin{center}

\tikzset{every picture/.style={line width=0.75pt}} %set default line width to 0.75pt        

\begin{tikzpicture}[x=0.75pt,y=0.75pt,yscale=-1,xscale=1]

%Curve Lines [id:da42227886447743124] 
\draw    (50.75,168.5) .. controls (-21.5,169.5) and (5.5,218.5) .. (50.75,213) ;
%Flowchart: Connector [id:dp833556052694024] 
\draw   (40,190.75) .. controls (40,178.46) and (44.81,168.5) .. (50.75,168.5) .. controls (56.69,168.5) and (61.5,178.46) .. (61.5,190.75) .. controls (61.5,203.04) and (56.69,213) .. (50.75,213) .. controls (44.81,213) and (40,203.04) .. (40,190.75) -- cycle ;
%Curve Lines [id:da11969617127657828] 
\draw [color={rgb, 255:red, 74; green, 144; blue, 226 }  ,draw opacity=1 ]   (52.54,213.92) .. controls (124.73,210.79) and (96.3,162.61) .. (51.23,169.44) ;
%Flowchart: Connector [id:dp542938163927856] 
\draw  [color={rgb, 255:red, 74; green, 144; blue, 226 }  ,draw opacity=1 ] (62.63,191.36) .. controls (62.99,203.65) and (58.47,213.75) .. (52.54,213.92) .. controls (46.61,214.1) and (41.5,204.28) .. (41.14,192) .. controls (40.78,179.71) and (45.29,169.61) .. (51.23,169.44) .. controls (57.16,169.26) and (62.27,179.08) .. (62.63,191.36) -- cycle ;

% Text Node
\draw (64,185.4) node [anchor=north west][inner sep=0.75pt]  [xscale=0.5,yscale=0.5]  {$\color{red}\bullet$};

\end{tikzpicture},
\end{center}
is not vanishing, i.e. $\tau({\color{red}\bullet})\neq0$. The first issue raised at the beginning of this section can be resolved by imposing the normalization condition $\tau({\color{red}\bullet})=1$. Moreover, the topological charge of the background particle exhibits genus dependence through the relationship: $q_g({\color{red}\bullet})\propto(1-g)$, where $g$ is the Riemann surface genus. This confirms that our background particle indeed encodes the theory's anomaly structure. For example, in the $\mathbb{CP}^{n}$ non-linear sigma model (NLSM), the explicit realization is:${\color{red}\bullet}\coloneqq  (\sigma/\Lambda)^{n(1-g)}$.

A critical reader might question the topological consistency of this construction, since the underlying puzzle remains not fully resolved: the cylinder topology comprises two circles, yet the composition of the co-multiplication and our modified trace map results in three circles (one of which is linked to the background particle). Moreover, it seemingly reproduces only conventional correlation functions derivable from the elementary bordisms in Section \ref{AS:axioms}, suggesting limited novelty.

Regarding the first point, we argued in the introduction that the conventional understanding of the topology of the trace map $\tau$ is misleading. While $\tau(1)$ would naively yield a sphere, the anomaly of the axial R-charge forces this result to vanish. This vanishing is remedied by the inclusion of background particles, a phenomenon that is clearly manifested in the associated 3D theory. As for the second point, although the vector space structures are topologically dictated by bordisms, the crucial innovation resides in the modified inner product. This new inner product is induced by our novel bordism configuration:

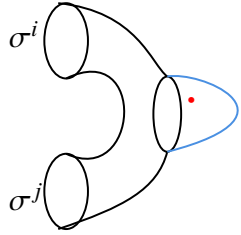
\begin{center}

\tikzset{every picture/.style={line width=0.75pt}} %set default line width to 0.75pt        

\begin{tikzpicture}[x=0.75pt,y=0.75pt,yscale=-1,xscale=1]

%Flowchart: Connector [id:dp9097939274941981] 
\draw   (33.5,159) .. controls (33.5,148.51) and (38.42,140) .. (44.5,140) .. controls (50.58,140) and (55.5,148.51) .. (55.5,159) .. controls (55.5,169.49) and (50.58,178) .. (44.5,178) .. controls (38.42,178) and (33.5,169.49) .. (33.5,159) -- cycle ;
%Flowchart: Connector [id:dp16205831416792082] 
\draw   (33.5,235) .. controls (33.5,224.51) and (38.42,216) .. (44.5,216) .. controls (50.58,216) and (55.5,224.51) .. (55.5,235) .. controls (55.5,245.49) and (50.58,254) .. (44.5,254) .. controls (38.42,254) and (33.5,245.49) .. (33.5,235) -- cycle ;
%Flowchart: Connector [id:dp03256465579509271] 
\draw   (88.5,196) .. controls (88.5,185.51) and (91.63,177) .. (95.5,177) .. controls (99.37,177) and (102.5,185.51) .. (102.5,196) .. controls (102.5,206.49) and (99.37,215) .. (95.5,215) .. controls (91.63,215) and (88.5,206.49) .. (88.5,196) -- cycle ;
%Curve Lines [id:da35909665547671676] 
\draw    (40.5,140) .. controls (80.5,144) and (87.5,171) .. (95.5,177) ;
%Curve Lines [id:da9705445499347057] 
\draw    (40.5,254) .. controls (73.5,243) and (90.5,234) .. (95.5,215) ;
%Curve Lines [id:da039086602323399666] 
\draw    (44.5,178) .. controls (85.5,158) and (81.5,233) .. (44.5,216) ;
%Curve Lines [id:da6759190596372209] 
\draw [color={rgb, 255:red, 74; green, 144; blue, 226 }  ,draw opacity=1 ]   (95.5,177) .. controls (104.06,175.57) and (120.31,180.03) .. (127.47,187.03) .. controls (135.37,194.75) and (132.21,205.56) .. (95.5,215) ;

% Text Node
\draw (14,147.4) node [anchor=north west][inner sep=0.75pt]  [xscale=1,yscale=1]  {$\sigma^i$};
% Text Node
\draw (14,229.4) node [anchor=north west][inner sep=0.75pt]  [anchor=north west][inner sep=0.75pt]  [xscale=1,yscale=1]  {$\sigma^j$};
% Text Node
\draw (105,186.4) node [anchor=north west][inner sep=0.75pt]  [xscale=0.5,yscale=0.5]  {$\color{red}\bullet$};

\end{tikzpicture}

\end{center}
giving
\begin{equation}\label{WIP:Funt}
    Q^{W}(\sigma^i, \sigma^j)\coloneqq \tau\left(\sigma^i\cdot\sigma^j \cdot{\color{red}\bullet}\right). 
\end{equation}
This bordism is constructed through the composition of the multiplication and co-unit maps. 

\noindent{\textbf{Unitarity in Witten-type TQFTs}}: We analyze unitarity through the concrete example of the A-twisted NLSM on $\mathbb{CP}^n$. The modified inner product in this case takes the diagonal form:
\begin{equation}\nonumber 
    g_{\bar{i}j}=\begin{pmatrix}
\Lambda^{-n} &  &\\
 & \ddots &\\
 &  & \Lambda^{-n} \\
\end{pmatrix}.
\end{equation}
For the theory to be unitary under this new inner product, we must require the dynamical scale $\Lambda$ to be strictly positive real.

Analogously, the composition of the co-multiplication and unit maps yields:

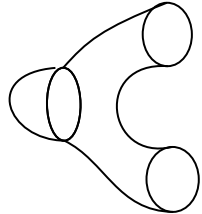
\begin{center}
    
\tikzset{every picture/.style={line width=0.75pt}} %set default line width to 0.75pt        

\begin{tikzpicture}[x=0.75pt,y=0.75pt,yscale=-1,xscale=1]

%Flowchart: Connector [id:dp08136170515797969] 
\draw   (39.28,225.71) .. controls (39.43,235.87) and (35.77,244.16) .. (31.1,244.23) .. controls (26.42,244.31) and (22.51,236.13) .. (22.35,225.97) .. controls (22.2,215.8) and (25.86,207.51) .. (30.53,207.44) .. controls (35.21,207.37) and (39.12,215.54) .. (39.28,225.71) -- cycle ;
%Flowchart: Connector [id:dp8108131303401963] 
\draw   (39.28,225.71) .. controls (39.43,235.87) and (35.77,244.16) .. (31.1,244.23) .. controls (26.42,244.31) and (22.51,236.13) .. (22.35,225.97) .. controls (22.2,215.8) and (25.86,207.51) .. (30.53,207.44) .. controls (35.21,207.37) and (39.12,215.54) .. (39.28,225.71) -- cycle ;
%Curve Lines [id:da28010410814631337] 
\draw    (31.1,244.23) .. controls (2.16,243.88) and (-10.96,210.48) .. (26.6,207.5) ;
%Flowchart: Connector [id:dp08610900734881333] 
\draw   (98.91,263.59) .. controls (99.05,272.65) and (93.26,280.08) .. (85.98,280.19) .. controls (78.7,280.31) and (72.68,273.06) .. (72.54,264) .. controls (72.4,254.94) and (78.19,247.51) .. (85.48,247.4) .. controls (92.76,247.29) and (98.78,254.54) .. (98.91,263.59) -- cycle ;
%Flowchart: Connector [id:dp2745743267163877] 
\draw   (95.43,190.44) .. controls (95.56,199.27) and (90.12,206.52) .. (83.28,206.63) .. controls (76.43,206.73) and (70.77,199.66) .. (70.63,190.82) .. controls (70.5,181.98) and (75.94,174.74) .. (82.78,174.63) .. controls (89.63,174.53) and (95.29,181.6) .. (95.43,190.44) -- cycle ;
%Curve Lines [id:da10659859126156224] 
\draw    (85.98,280.19) .. controls (57.63,279.83) and (51.92,254.31) .. (31.1,244.23) ;
%Curve Lines [id:da19407929363006826] 
\draw    (82.78,174.63) .. controls (49.33,187.95) and (42.5,194) .. (30.53,207.44) ;
%Curve Lines [id:da16166389595555453] 
\draw    (85.48,247.4) .. controls (51.93,255.11) and (45.64,204.01) .. (83.28,206.63) ;

\end{tikzpicture}
\end{center}

\begin{equation}\label{CWIP:Funt}
    \Delta^{W}: {\mathbb{C}}\rightarrow \sum_{ij}\Delta^{W}_{ij}\sigma^i\otimes \sigma^j. 
\end{equation}
One immediate consistency check: the new S-diagram shall be isomorphic to the cylinder map that can be easily seen as following
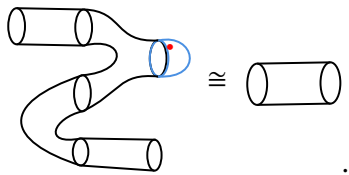
\begin{center}

\tikzset{every picture/.style={line width=0.75pt}} %set default line width to 0.75pt        

\begin{tikzpicture}[x=0.75pt,y=0.75pt,yscale=-1,xscale=1]

%Flowchart: Connector [id:dp011063192371274178] 
\draw   (0,194.11) .. controls (0,189.08) and (1.89,185) .. (4.23,185) .. controls (6.56,185) and (8.45,189.08) .. (8.45,194.11) .. controls (8.45,199.14) and (6.56,203.22) .. (4.23,203.22) .. controls (1.89,203.22) and (0,199.14) .. (0,194.11) -- cycle ;
%Flowchart: Connector [id:dp8174486810427456] 
\draw   (33.81,194.76) .. controls (33.81,189.73) and (35.71,185.65) .. (38.04,185.65) .. controls (40.37,185.65) and (42.27,189.73) .. (42.27,194.76) .. controls (42.27,199.79) and (40.37,203.87) .. (38.04,203.87) .. controls (35.71,203.87) and (33.81,199.79) .. (33.81,194.76) -- cycle ;
%Flowchart: Connector [id:dp9862811705361673] 
\draw   (33.19,227.95) .. controls (33.19,222.92) and (35.08,218.84) .. (37.41,218.84) .. controls (39.75,218.84) and (41.64,222.92) .. (41.64,227.95) .. controls (41.64,232.98) and (39.75,237.06) .. (37.41,237.06) .. controls (35.08,237.06) and (33.19,232.98) .. (33.19,227.95) -- cycle ;
%Flowchart: Connector [id:dp805995550110916] 
\draw   (32.56,257.56) .. controls (32.56,253.79) and (34.31,250.73) .. (36.47,250.73) .. controls (38.64,250.73) and (40.39,253.79) .. (40.39,257.56) .. controls (40.39,261.34) and (38.64,264.4) .. (36.47,264.4) .. controls (34.31,264.4) and (32.56,261.34) .. (32.56,257.56) -- cycle ;
%Flowchart: Connector [id:dp8153354490385744] 
\draw   (70.13,259.19) .. controls (70.13,254.88) and (71.74,251.38) .. (73.73,251.38) .. controls (75.72,251.38) and (77.33,254.88) .. (77.33,259.19) .. controls (77.33,263.5) and (75.72,267) .. (73.73,267) .. controls (71.74,267) and (70.13,263.5) .. (70.13,259.19) -- cycle ;
%Flowchart: Connector [id:dp6853320000252617] 
\draw   (71.38,210.38) .. controls (71.38,205.35) and (73.28,201.27) .. (75.61,201.27) .. controls (77.94,201.27) and (79.84,205.35) .. (79.84,210.38) .. controls (79.84,215.41) and (77.94,219.49) .. (75.61,219.49) .. controls (73.28,219.49) and (71.38,215.41) .. (71.38,210.38) -- cycle ;
%Flowchart: Connector [id:dp519795747555411] 
\draw  [color={rgb, 255:red, 74; green, 144; blue, 226 }  ,draw opacity=1 ] (71.38,210.38) .. controls (71.38,205.35) and (73.49,201.27) .. (76.08,201.27) .. controls (78.67,201.27) and (80.78,205.35) .. (80.78,210.38) .. controls (80.78,215.41) and (78.67,219.49) .. (76.08,219.49) .. controls (73.49,219.49) and (71.38,215.41) .. (71.38,210.38) -- cycle ;
%Straight Lines [id:da6688821473729566] 
\draw    (4.23,185) -- (38.04,185.65) ;
%Straight Lines [id:da350229110057259] 
\draw    (4.23,203.22) -- (38.04,203.87) ;
%Curve Lines [id:da7591223853929172] 
\draw    (38.04,185.65) .. controls (58.55,186.3) and (52.91,202.57) .. (75.61,201.27) ;
%Curve Lines [id:da41209042603670143] 
\draw    (37.41,237.06) .. controls (56.67,227.95) and (55.42,217.54) .. (75.61,219.49) ;
%Curve Lines [id:da30295384028613426] 
\draw    (38.04,203.87) .. controls (59.17,198.67) and (61.68,216.89) .. (37.41,218.84) ;
%Curve Lines [id:da5126720363790401] 
\draw [color={rgb, 255:red, 74; green, 144; blue, 226 }  ,draw opacity=1 ]   (76.08,201.27) .. controls (94.24,197.37) and (99.25,220.79) .. (75.61,219.49) ;
%Curve Lines [id:da07049637586967938] 
\draw    (37.41,218.84) .. controls (-5.32,233.16) and (-1.88,252.03) .. (36.47,264.4) ;
%Curve Lines [id:da545042461195325] 
\draw    (37.41,237.06) .. controls (22.86,240.32) and (16.59,254.63) .. (36.79,250.73) ;
%Straight Lines [id:da22529724913863347] 
\draw    (36.79,250.73) -- (74.36,251.38) ;
%Straight Lines [id:da41622149613219894] 
\draw    (36.47,264.4) -- (73.73,267) ;
%Flowchart: Connector [id:dp8225206493338337] 
\draw   (120.54,223.4) .. controls (120.54,217.65) and (122.78,212.98) .. (125.55,212.98) .. controls (128.31,212.98) and (130.56,217.65) .. (130.56,223.4) .. controls (130.56,229.15) and (128.31,233.81) .. (125.55,233.81) .. controls (122.78,233.81) and (120.54,229.15) .. (120.54,223.4) -- cycle ;
%Flowchart: Connector [id:dp5724020207680944] 
\draw   (157.48,222.75) .. controls (157.48,217) and (159.72,212.33) .. (162.49,212.33) .. controls (165.26,212.33) and (167.5,217) .. (167.5,222.75) .. controls (167.5,228.5) and (165.26,233.16) .. (162.49,233.16) .. controls (159.72,233.16) and (157.48,228.5) .. (157.48,222.75) -- cycle ;
%Straight Lines [id:da7389913339188692] 
\draw    (125.55,212.98) -- (162.49,212.33) ;
%Straight Lines [id:da5495311637276341] 
\draw    (162.49,233.16) -- (125.55,233.81) ;

% Text Node
\draw (78.85,202.13) node [anchor=north west][inner sep=0.75pt]  [xscale=0.5,yscale=0.5]  {$\color{red}\bullet$};
% Text Node
\draw (98.26,215.8) node [anchor=north west][inner sep=0.75pt]  [xscale=1,yscale=1]  {$\cong$};
\end{tikzpicture}.
\end{center}

Furthermore, we must examine how orientation reversal and duality transformations affect the elementary bordisms. For the original bordisms in Section \ref{AS:axioms}, we immediately obtain:  
\begin{equation*}
    \bar{m}^{\vee}=\Delta, \bar{\Delta}^{\vee}=m, \bar{1}^{\vee}=\tau, \bar{\tau}^{\vee}=1. 
\end{equation*}
However, in our extended framework with background insertions, the last two relations require modification:
\begin{center}
\tikzset{every picture/.style={line width=0.75pt}} %set default line width to 0.75pt        

\begin{tikzpicture}[x=0.75pt,y=0.75pt,yscale=-1,xscale=1]

%Flowchart: Connector [id:dp8300796298615074] 
\draw  [color={rgb, 255:red, 74; green, 144; blue, 226 }  ,draw opacity=1 ] (92.3,196.4) .. controls (92.3,187.12) and (93.77,179.6) .. (95.59,179.6) .. controls (97.4,179.6) and (98.87,187.12) .. (98.87,196.4) .. controls (98.87,205.68) and (97.4,213.2) .. (95.59,213.2) .. controls (93.77,213.2) and (92.3,205.68) .. (92.3,196.4) -- cycle ;
%Curve Lines [id:da29493658514573695] 
\draw [color={rgb, 255:red, 74; green, 144; blue, 226 }  ,draw opacity=1 ]   (95.59,179.6) .. controls (119.76,174.8) and (136.78,208.4) .. (95.59,213.2) ;
%Flowchart: Connector [id:dp8380399666143437] 
\draw   (37.38,168.4) .. controls (37.38,160.01) and (40.41,153.2) .. (44.15,153.2) .. controls (47.88,153.2) and (50.91,160.01) .. (50.91,168.4) .. controls (50.91,176.8) and (47.88,183.6) .. (44.15,183.6) .. controls (40.41,183.6) and (37.38,176.8) .. (37.38,168.4) -- cycle ;
%Flowchart: Connector [id:dp39264354232466625] 
\draw   (36.6,226.8) .. controls (36.6,218.41) and (39.98,211.6) .. (44.15,211.6) .. controls (48.31,211.6) and (51.69,218.41) .. (51.69,226.8) .. controls (51.69,235.19) and (48.31,242) .. (44.15,242) .. controls (39.98,242) and (36.6,235.19) .. (36.6,226.8) -- cycle ;
%Flowchart: Connector [id:dp12472117317437781] 
\draw  [color={rgb, 255:red, 0; green, 0; blue, 0 }  ,draw opacity=1 ] (92.3,196.4) .. controls (92.3,187.12) and (93.77,179.6) .. (95.59,179.6) .. controls (97.4,179.6) and (98.87,187.12) .. (98.87,196.4) .. controls (98.87,205.68) and (97.4,213.2) .. (95.59,213.2) .. controls (93.77,213.2) and (92.3,205.68) .. (92.3,196.4) -- cycle ;
%Curve Lines [id:da7801827532482821] 
\draw    (44.15,153.2) .. controls (74.89,146) and (70.25,173.2) .. (95.59,179.6) ;
%Curve Lines [id:da5522276668728876] 
\draw    (44.15,242) .. controls (75.09,218) and (64.64,237.2) .. (95.59,213.2) ;
%Curve Lines [id:da3041303120566555] 
\draw    (44.15,183.6) .. controls (75.09,159.6) and (105.84,218) .. (44.15,211.6) ;
%Flowchart: Connector [id:dp32607399680378324] 
\draw  [color={rgb, 255:red, 74; green, 144; blue, 226 }  ,draw opacity=1 ] (241.5,172.66) .. controls (241.41,180.94) and (239.78,187.63) .. (237.86,187.62) .. controls (235.94,187.6) and (234.44,180.89) .. (234.53,172.61) .. controls (234.61,164.34) and (236.24,157.64) .. (238.16,157.65) .. controls (240.09,157.67) and (241.58,164.39) .. (241.5,172.66) -- cycle ;
%Curve Lines [id:da035764665377266525] 
\draw [color={rgb, 255:red, 0; green, 0; blue, 0 }  ,draw opacity=1 ]   (237.86,187.62) .. controls (212.19,191.71) and (194.45,161.62) .. (238.16,157.65) ;
%Flowchart: Connector [id:dp36678243043402337] 
\draw   (234.71,172.84) .. controls (234.75,164.44) and (236.29,157.64) .. (238.16,157.65) .. controls (240.03,157.66) and (241.51,164.48) .. (241.47,172.87) .. controls (241.43,181.26) and (239.88,188.06) .. (238.02,188.05) .. controls (236.15,188.04) and (234.66,181.23) .. (234.71,172.84) -- cycle ;
%Flowchart: Connector [id:dp2667490320949061] 
\draw   (230.5,228.41) .. controls (230.53,222.03) and (232.58,216.87) .. (235.08,216.88) .. controls (237.58,216.89) and (239.59,222.07) .. (239.56,228.45) .. controls (239.53,234.83) and (237.47,239.99) .. (234.97,239.98) .. controls (232.47,239.97) and (230.47,234.79) .. (230.5,228.41) -- cycle ;
%Flowchart: Connector [id:dp6303046610982477] 
\draw  [color={rgb, 255:red, 0; green, 0; blue, 0 }  ,draw opacity=1 ] (241.5,172.66) .. controls (241.41,180.94) and (239.78,187.63) .. (237.86,187.62) .. controls (235.94,187.6) and (234.44,180.89) .. (234.53,172.61) .. controls (234.61,164.34) and (236.24,157.64) .. (238.16,157.65) .. controls (240.09,157.67) and (241.58,164.39) .. (241.5,172.66) -- cycle ;
%Curve Lines [id:da3538385329435737] 
\draw    (238.16,157.65) .. controls (268.95,150.6) and (266.2,184.48) .. (291.5,191) ;
%Curve Lines [id:da9967541717446263] 
\draw    (234.97,239.98) .. controls (266.03,216.13) and (260.44,237.85) .. (291.5,214) ;
%Curve Lines [id:da565518523326105] 
\draw    (238.23,188.9) .. controls (269.29,165.05) and (296.74,223.58) .. (235.08,216.88) ;
%Flowchart: Connector [id:dp550156334838902] 
\draw   (284.5,202.5) .. controls (284.5,196.15) and (287.63,191) .. (291.5,191) .. controls (295.37,191) and (298.5,196.15) .. (298.5,202.5) .. controls (298.5,208.85) and (295.37,214) .. (291.5,214) .. controls (287.63,214) and (284.5,208.85) .. (284.5,202.5) -- cycle ;

% Text Node
\draw (30.41,161.2) node [anchor=north west][inner sep=0.75pt]  [xscale=0.5,yscale=0.5]  {$\color{blue} \bullet$};
% Text Node
\draw (101,187.2) node [anchor=north west][inner sep=0.75pt]  [xscale=0.5,yscale=0.5]  {$\color{red} \bullet$};
% Text Node
\draw (33,255.4) node [anchor=north west][inner sep=0.75pt]  [xscale=1,yscale=1]  {$\bar{1}^{\vee}=\tau\circ m({\color{red} \bullet}\cdot{\color{blue} \bullet}=1, \quad)$};
% Text Node
\draw (225,251.4) node [anchor=north west][inner sep=0.75pt]  [xscale=1,yscale=1]  {$\bar{\tau}^{\vee}=m(1,{\color{blue} \bullet})$};
% Text Node
\draw (224,220.4) node [anchor=north west][inner sep=0.75pt]  [xscale=0.5,yscale=0.5]  {$\color{blue} \bullet$};

\end{tikzpicture},
\end{center}
where ${\color{blue}\bullet}$ denotes the background anti-particle satisfying the relation  ${\color{red} \bullet}\cdot{\color{blue} \bullet}=1$. These modified duality relations confirm that our bordism category admits a consistent unitary structure.

The bordism corresponding to the two-sphere partition function in Section \ref{AS:axioms} is now naturally generalizes to a correlation function in our extended framework:
\begin{center}

\tikzset{every picture/.style={line width=0.75pt}} %set default line width to 0.75pt        

\begin{tikzpicture}[x=0.75pt,y=0.75pt,yscale=-1,xscale=1]

%Flowchart: Connector [id:dp9097939274941981] 
\draw   (33.5,159) .. controls (33.5,148.51) and (38.42,140) .. (44.5,140) .. controls (50.58,140) and (55.5,148.51) .. (55.5,159) .. controls (55.5,169.49) and (50.58,178) .. (44.5,178) .. controls (38.42,178) and (33.5,169.49) .. (33.5,159) -- cycle ;
%Flowchart: Connector [id:dp16205831416792082] 
\draw   (33.5,235) .. controls (33.5,224.51) and (38.42,216) .. (44.5,216) .. controls (50.58,216) and (55.5,224.51) .. (55.5,235) .. controls (55.5,245.49) and (50.58,254) .. (44.5,254) .. controls (38.42,254) and (33.5,245.49) .. (33.5,235) -- cycle ;
%Flowchart: Connector [id:dp03256465579509271] 
\draw   (88.5,196) .. controls (88.5,185.51) and (91.63,177) .. (95.5,177) .. controls (99.37,177) and (102.5,185.51) .. (102.5,196) .. controls (102.5,206.49) and (99.37,215) .. (95.5,215) .. controls (91.63,215) and (88.5,206.49) .. (88.5,196) -- cycle ;
%Curve Lines [id:da35909665547671676] 
\draw    (40.5,140) .. controls (80.5,144) and (87.5,171) .. (95.5,177) ;
%Curve Lines [id:da9705445499347057] 
\draw    (40.5,254) .. controls (73.5,243) and (90.5,234) .. (95.5,215) ;
%Curve Lines [id:da039086602323399666] 
\draw    (44.5,178) .. controls (85.5,158) and (81.5,233) .. (44.5,216) ;
%Curve Lines [id:da6759190596372209] 
\draw [color={rgb, 255:red, 74; green, 144; blue, 226 }  ,draw opacity=1 ]   (95.5,177) .. controls (104.06,175.57) and (120.31,180.03) .. (127.47,187.03) .. controls (135.37,194.75) and (132.21,205.56) .. (95.5,215) ;
%Curve Lines [id:da5322736586647357] 
\draw    (44.5,216) .. controls (4.5,226) and (10.5,246) .. (44.5,254) ;
%Flowchart: Connector [id:dp9850528133423123] 
\draw  [color={rgb, 255:red, 74; green, 144; blue, 226 }  ,draw opacity=1 ] (181,191) .. controls (181,183.82) and (183.8,178) .. (187.25,178) .. controls (190.7,178) and (193.5,183.82) .. (193.5,191) .. controls (193.5,198.18) and (190.7,204) .. (187.25,204) .. controls (183.8,204) and (181,198.18) .. (181,191) -- cycle ;
%Flowchart: Connector [id:dp34941803463028176] 
\draw   (181.5,192) .. controls (181.5,184.82) and (184.3,179) .. (187.75,179) .. controls (191.2,179) and (194,184.82) .. (194,192) .. controls (194,199.18) and (191.2,205) .. (187.75,205) .. controls (184.3,205) and (181.5,199.18) .. (181.5,192) -- cycle ;
%Curve Lines [id:da8868437371696202] 
\draw    (187.75,179) .. controls (156.5,181) and (165.5,206) .. (187.75,205) ;
%Curve Lines [id:da8792459089409034] 
\draw [color={rgb, 255:red, 74; green, 144; blue, 226 }  ,draw opacity=1 ]   (187.75,179) .. controls (205,174) and (218,199) .. (187.75,205) ;

% Text Node
\draw (28,155.4) node [anchor=north west][inner sep=0.75pt]  [xscale=0.5,yscale=0.5]  {$\color{blue}\bullet$};
% Text Node
\draw (105,186.4) node [anchor=north west][inner sep=0.75pt]  [xscale=0.5,yscale=0.5]  {$\color{red}\bullet$};
% Text Node
\draw (140,185.4) node [anchor=north west][inner sep=0.75pt]  [xscale=1,yscale=1]  {$\cong$};
\end{tikzpicture},
\end{center}
which is vanishing: $\tau({\color{red}\bullet}\cdot{\color{blue}\bullet}=1)=0$. 

Building upon these inputs and results, we formulate the following conjecture  (structure theorem for Witten-type TQFT bordisms):
\begin{center}
\textit{Every bordism in a Witten-type TQFT admits a canonical decomposition into the elementary building blocks defined in this section, while preserving all algebraic relations established in Section \ref{AS:axioms}.}
\end{center}
Furthermore, for any bordism with multiple distinct decompositions, there exists a sewing theorem that guarantees consistency between different decompositions without requiring additional relations beyond those specified in Section \ref{AS:axioms}.} 

An intuitive justification for this conjecture emerges through a correspondence between the trace structures in our framework and those in Section  \ref{AS:axioms}:
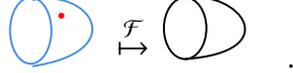
\begin{center}

\tikzset{every picture/.style={line width=0.75pt}} %set default line width to 0.75pt        

\begin{tikzpicture}[x=0.75pt,y=0.75pt,yscale=-1,xscale=1]

%Flowchart: Connector [id:dp9850528133423123] 
\draw  [color={rgb, 255:red, 74; green, 144; blue, 226 }  ,draw opacity=1 ] (3,223.01) .. controls (3,213.78) and (7.67,206.3) .. (13.43,206.3) .. controls (19.19,206.3) and (23.86,213.78) .. (23.86,223.01) .. controls (23.86,232.23) and (19.19,239.71) .. (13.43,239.71) .. controls (7.67,239.71) and (3,232.23) .. (3,223.01) -- cycle ;
%Curve Lines [id:da8792459089409034] 
\draw [color={rgb, 255:red, 74; green, 144; blue, 226 }  ,draw opacity=1 ]   (14.26,207.58) .. controls (43.05,201.16) and (64.74,233.29) .. (14.26,241) ;
%Flowchart: Connector [id:dp979730245335008] 
\draw   (80.58,221.82) .. controls (80.58,213.25) and (85.38,206.3) .. (91.3,206.3) .. controls (97.22,206.3) and (102.02,213.25) .. (102.02,221.82) .. controls (102.02,230.4) and (97.22,237.35) .. (91.3,237.35) .. controls (85.38,237.35) and (80.58,230.4) .. (80.58,221.82) -- cycle ;
%Curve Lines [id:da46387067315982733] 
\draw    (91.3,206.3) .. controls (119.37,202.65) and (142.18,233.69) .. (91.3,237.35) ;

% Text Node
\draw (26.21,213.31) node [anchor=north west][inner sep=0.75pt]  [xscale=0.5,yscale=0.5]  {${\color{red}\bullet}$};
% Text Node
\draw (55.41,214.22) node [anchor=north west][inner sep=0.75pt]  [xscale=1,yscale=1]  {$\overset{\cal F}\mapsto$};

\end{tikzpicture}.
\end{center}
All bordisms constructed from our elementary generators are mapped to their counterparts in the Schwarz-type TQFTs of Section \ref{AS:axioms}. The established theorems for Schwarz-type TQFTs then become applicable. A crucial subtlety arises under orientation reversal: the mapping transforms non-trivially as
\begin{equation}
    \overline{{\cal F}(\tau)}^{\vee} \neq {\cal F}(\bar{\tau}^{\vee})
\end{equation}
demonstrating that the mapping cannot be realized as a simple linear transformation.

\section{Summing over bordisms in TQFT}\label{SBTQFT}

Our bordism definitions might suggest that semi-simple Witten-type TQFTs are identical to their associated Schwarz-type counterparts, however, this section reveals a crucial distinction.

As emphasized in \cite{Banerjee:2022pmw}, the summation over all bordisms to compute amplitudes (unlike in quantum gravity) constitutes a non-trivial operation unique to the TQFT framework. This stems from the fundamental property that all operators acting on the state space $\cal C$ are topologically encoded by bordisms. Specifically, for the vacuum-to-vacuum amplitude in the 2D theories of Section \ref{AS:axioms}, this bordism summation for connected surfaces reduces to:
\begin{equation*}
    Z=Z(S^2)+Z(T^2)+\cdots.
\end{equation*}
In \cite{Banerjee:2022pmw}, it was shown that the total amplitude, when including disconnected surfaces, is given by
\begin{equation*}
   {\cal A}= \exp{Z}
\end{equation*}
At the operator level, the symmetric monoidal functor implies the following:
\begin{equation*}
Z=\tau(1+{\cal H}+\cdots)=\tau\left(\frac{1}{1-\cal H}\right),
\end{equation*}
where we have formally summed the operator series. This demonstrates:
\begin{center}
the topological summation in 2D TQFTs is mathematically equivalent to requiring the existence of the resolvent operator $({1-\cal H})^{-1}$ in the vector space. 
\end{center}
The existence of the resolvent operator $({1-\cal H})^{-1}$ is guaranteed when the spectral radius of $\cal H$ is smaller than one. For example, for the Verlinde algebra of $U(1)_n$, its handle operator possesses a unit eigenvalue, which precludes a convergent topological summation. In contrast, the Verlinde algebra of $SU(k)_n$ admits such a summation, as its handle operator spectrum satisfies the convergence criterion.

The topological summation is rigorously defined only within a single super-selection sector. While all bordisms in the previous case reside in the same sector, the situation becomes more subtle for Witten-type TQFTs (as analyzed in Section \ref{BWTQFT}), where bordisms may span multiple sectors.

This distinction originates from the anomaly dependence of super-selection sectors. Specifically, for the A-twisted NLSM on a K\"ahler target $\cal M$, the anomaly index for degree-$d$ maps is given by:

\begin{equation*}
  A=  (1-g)\dim{\cal M}+d\cdot c_1(\cal M).
\end{equation*}
where $c_1(\cal M)$ denotes the first Chern class. Crucially, vacuum-to-vacuum amplitudes require summing solely over bordisms with $A=0$, a condition that cannot be satisfied for all genera for the connected surface, marking a fundamental departure from the scenario in Section \ref{AS:axioms}. 

We conclude this section by analyzing the canonical example of $\mathbb{CP}^n$. The anomaly index is given explicitly by $A=n(1-g)+d(1+n)$, which restricts the topological sum to connected surfaces of genus $g=l(1+n)+1$ for integer $l\geq 0$. The corresponding operator has the following structure:
\begin{equation}\label{WTGH}
 \frac{{\cal H}}{1-{\cal H}^{1+n}}.
\end{equation}
For each genus $g$, the partition function admits the following expression \cite{Vafa:1990mu, Gu:2020nub}:
\begin{equation*}
 \sum_{\sigma_0} (n+1)^{g-1}\sigma^{n(g-1)}, \quad \sigma_0\in \{\sigma^{n+1}=q\},
\end{equation*}
where for the admissible genus $g=l(1+n)+1$, the factor $\sigma^{n(g-1)}$ simplifies to $q^{nl}$. The convergence condition for (\ref{WTGH}) requires the constraint: $q<(n+1)^{-\left(1+\frac{2}{n}\right)}$.

\section*{Acknowledgements}
We thank Gregory Moore and Du Pei for useful discussions.

\end{document}